\documentclass[aps,pra,twocolumn,showpacs,amsmath,amssymb,superscriptaddress]{revtex4-1}

\usepackage{graphicx, amsmath, bm, braket, txfonts, color, ulem}

\usepackage{mathrsfs}

\usepackage{amssymb}

\usepackage[colorlinks=true, citecolor=blue, filecolor=blue, linkcolor=blue, urlcolor=blue]{hyperref}

\begin{document}

\title{Low-symmetry nanowire cross-sections for enhanced Dresselhaus spin-orbit interaction}

\author{Miguel J. Carballido}
\affiliation{Department of Physics, University of Basel, Klingelbergstrasse 82, CH-4056 Basel, Switzerland}
\author{Christoph Kloeffel}
\affiliation{Department of Physics, University of Basel, Klingelbergstrasse 82, CH-4056 Basel, Switzerland}
\author{Dominik M. Zumb\"uhl}
\affiliation{Department of Physics, University of Basel, Klingelbergstrasse 82, CH-4056 Basel, Switzerland}
\author{Daniel Loss}
\affiliation{Department of Physics, University of Basel, Klingelbergstrasse 82, CH-4056 Basel, Switzerland}
\affiliation{CEMS, RIKEN, Wako, Saitama 351-0198, Japan}

\date{\today}

\begin{abstract}
We study theoretically the spin-orbit interaction of low-energy electrons in semiconducting nanowires with a zinc-blende lattice. The effective Dresselhaus term is derived for various growth directions, including $\langle 11\bar{2} \rangle$-oriented nanowires. While a specific configuration exists where the Dresselhaus spin-orbit coupling is suppressed even at confinement potentials of low symmetry, many configurations allow for a strong Dresselhaus coupling. In particular, we discuss qualitative and quantitative results for nanowire cross-sections modeled after sectors of rings or circles. The parameter dependence is analyzed in detail, enabling predictions for a large variety of setups. For example, we gain insight into the spin-orbit coupling in recently fabricated GaAs-InAs nanomembrane-nanowire structures. By combining the effective Dresselhaus and Rashba terms, we find that such structures are promising platforms for applications where an electrically controllable spin-orbit interaction is needed. If the nanowire cross-section is scaled down and InAs replaced by InSb, remarkably high Dresselhaus-based spin-orbit energies of the order of millielectronvolt are expected. A Rashba term that is similar to the effective Dresselhaus term can be induced via electric gates, providing means to switch the spin-orbit interaction on and off. By varying the central angle of the circular sector, we find, among other things, that particularly strong Dresselhaus couplings are possible when nanowire cross-sections resemble half-disks. 
\end{abstract}

\maketitle

\section{Introduction}

Semiconducting nanowires (NWs) are currently among the most promising building blocks for a large-scale, solid-state quantum computer. In particular, they may allow not only for conventional spin \cite{loss:pra98, nadjperge:nat10, petersson:nat12, kloeffel:annurev13, maurand:ncomm16} and charge \cite{wang:nlt19} qubits but also for topological quantum computing \cite{alicea:rpp12, beenakker:annurev13, klinovaja:prl14, klinovaja:prb14, lutchyn:nrm18}. The proposed schemes usually rely on spin-orbit interaction (SOI), which is a crucial mechanism in modern fields of condensed matter physics \cite{manchon:nmat15}.  

For an electron with spin $\bm{s}$ and momentum $\hbar \bm{k}$, the SOI can be considered as a coupling term proportional to $\bm{b}_{\rm SO}(\bm{k}) \cdot \bm{s}$, where $\bm{b}_{\rm SO}(\bm{k})$ is an effective magnetic field that depends on the momentum \cite{nadjperge:prl12}. Suitable setup geometries in experiments are often determined by the orientation of $\bm{b}_{\rm SO}(\bm{k})$. For example, electric dipole spin resonance is efficient when the externally applied magnetic field $\bm{B}$ is perpendicular to the effective field caused by SOI \cite{rashba:prl03, golovach:prb06edsr, flindt:prl06}. Furthermore, the special geometry $\bm{B} \perp \bm{b}_{\rm SO}(\bm{k})$ is assumed in proposals for realizing Majorana fermions in NWs with proximity-induced superconductivity \cite{lutchyn:prl10, oreg:prl10}. Profound knowledge of $\bm{b}_{\rm SO}(\bm{k})$ is therefore essential. A prominent contribution to the SOI of electrons is the Rashba spin-orbit interaction (RSOI) \cite{bihlmayer:njp15, bychkov:jetp84, bychkov:jpcssp84, winkler:book}, which results from structure inversion asymmetry and can be controlled to a great extent by applying electric fields \cite{nitta:prl97, engels:prb97, liang:nl12, weigele:arX18}. The Dresselhaus spin-orbit interaction (DSOI) \cite{dresselhaus:pr55, winkler:book, hanson:rmp07}, which arises from an inversion asymmetry of the underlying crystal structure, is an equally important contribution. The effective DSOI term depends strongly on details of the electron confinement and, moreover, on the orientation of the crystallographic axes \cite{dresselhaus:pr55, winkler:book, hanson:rmp07, dyakonov:sps86, balocchi:prl11, flatte:physics11, luo:prb11, ganichev:pssb14, kammermeier:prb16, campos:prb18}. This holds true not only for two-dimensional (2D) systems, like quantum wells and lateral quantum dots, but also for one-dimensional (1D) systems like NWs. If the growth direction, the quantum confinement, and the applied electric fields are chosen appropriately, the Rashba and Dresselhaus contributions can result in a large $\bm{b}_{\rm SO}(\bm{k})$ or even cancel each other, at least in good approximation \cite{balocchi:prl11, flatte:physics11}, which can be used to switch the SOI on and off. The interplay between Dresselhaus and Rashba coupling also provides means to implement spin field-effect transistors in 2D and 1D devices that can operate in a nonballistic (or diffusive) regime \cite{schliemann:prl03}.     

Semiconducting NWs have been fabricated for several decades \cite{wagner:apl64, guniat:cr19}. Their cross-sections depend on details of the fabrication process. By now, a remarkable variety of cross-sections has been reported, ranging from approximately circular \cite{casse:apl10, barraud:edl12}, hexagonal \cite{wu:nl04, fortuna:sst10, hauge:nl15, takase:scirep17}, or (with various aspect ratios) rectangular \cite{coquand:ulisproc12, voisin:nlt16, calahorra:scirep17} to very special shapes. Germanium hut wires, for instance, are available since 2012 \cite{zhang:prl12} and attracted wide interest \cite{watzinger:nlt16, li:nl18, watzinger:ncomm18}. Their cross-section resembles an obtuse isosceles triangle. Very recently, Friedl \textit{et al.}~\cite{friedl:nl18} reported the template-assisted growth of InAs NW networks. A striking feature of these scalable networks is the demonstrated possibility to create Y-shaped NW junctions. Such junctions are useful, e.g., for braiding Majorana fermions \cite{alicea:nphys11, beenakker:annurev13}. Since the NWs of Ref.~\cite{friedl:nl18} form on nanomembranes, their cross-section resembles a (major) circular sector. Due to this unusual NW cross-section, detailed information about the associated SOI is desirable. 

In this paper, we theoretically study the effective DSOI and RSOI of low-energy electrons in NWs with low-symmetry cross-sections. In particular, we consider cross-sections which are circular sectors of arbitrary central angle. Furthermore, we allow for a nonzero inner radius and analyze how the SOI depends on, e.g., the inner and outer radius, the sectorial angle, and the orientation of the main crystallographic axes. In agreement with previous calculations, which were performed for $\langle 100 \rangle$, $\langle 110 \rangle$, and $\langle 111 \rangle$-oriented NWs \cite{luo:prb11, kammermeier:prb16, campos:prb18, bringer:prb19}, we find that the growth direction affects the effective DSOI significantly. These earlier works are extended here by studying novel cross-sections and, moreover, by including $\langle 11\bar{2} \rangle$-oriented NWs. As recently demonstrated by Friedl \textit{et al.}~\cite{friedl:nl18} and Aseev \textit{et al.}~\cite{aseev:nl19}, $\langle 11\bar{2} \rangle$-oriented NWs can now be used to fabricate scalable NW networks whose NW junctions may, for instance, enable topological quantum information processing. We believe that our qualitative and quantitative results will allow everyone to quickly obtain reasonable estimates of the spin-orbit coupling, the effective magnetic field $\bm{b}_{\rm SO}$, the spin-orbit length, and the spin-orbit energy in various NWs. Given the NWs of Ref.~\cite{friedl:nl18}, for example, we expect $\bm{b}_{\rm SO}$ to be parallel to the substrate and perpendicular to the NW. If the cross-section of these NWs can be scaled down and if InAs can be replaced by InSb \cite{nadjperge:prl12, mourik:sci12, vandenberg:prl13, vanweperen:prb15, aseev:nl19}, which has a narrower band gap and a much larger Dresselhaus coefficient \cite{winkler:book, gmitra:prb16}, we find that the spin-orbit energy can exceed one millielectronvolt even without applied electric fields and, remarkably, that the effective SOI can be tuned continuously and switched on/off (apart from corrections which are cubic in the momentum) by applying an electric field perpendicular to the substrate. We also find, among other things, that a particularly strong DSOI is achievable with NWs whose cross-sections resemble half-disks. However, as we show here, a specific orientation of the crystallographic axes exists with which the effective DSOI is strongly suppressed for all considered cross-sections and despite their low symmetry.   

The paper is organized as follows. In Sec.~\ref{sec:NWCrossSectionsBasisStates}, we discuss the considered NW cross-sections and explain our calculation of the eigenstates in the absence of SOI. The effective DSOI term is obtained qualitatively in Sec.~\ref{sec:CalculationEffDSOI} and quantitatively in Sec.~\ref{sec:NumericalResults}. The effective RSOI term is studied in Sec.~\ref{sec:EffRashbaSOI}, followed by a concluding discussion in Sec.~\ref{sec:Discussion}. The appendix provides the details of the theory and, among other things, shows the effective DSOI terms for commonly used NW growth directions.

\section{Nanowire cross-sections and basis states}
\label{sec:NWCrossSectionsBasisStates}

\subsection{Hard-wall confinement}
\label{secsub:HardWallConf} 

\begin{figure}[htb]
\centering
\includegraphics[width=0.75\linewidth]{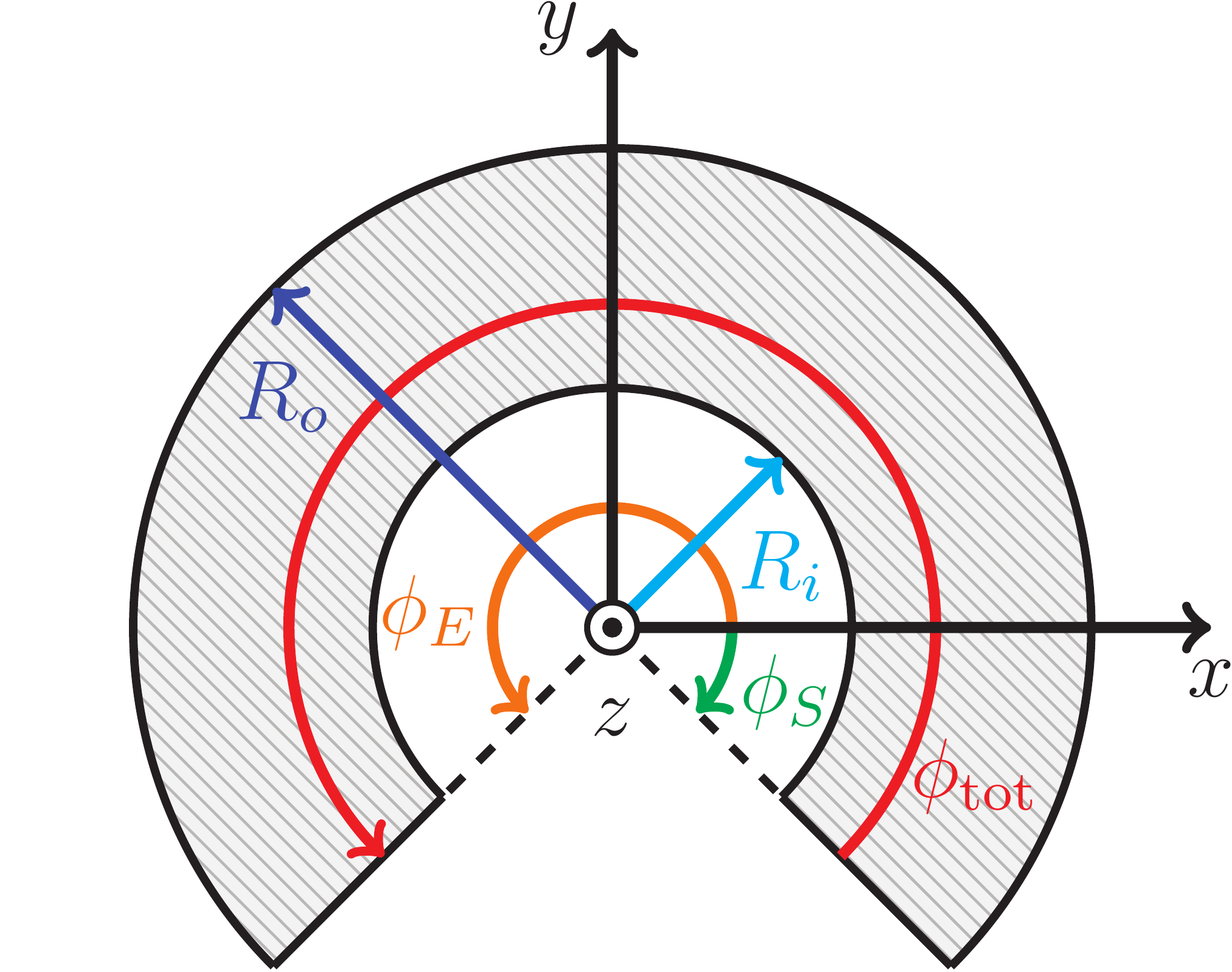} 
\caption{Generic sketch of the sectorial annular cross-section (SAC) used to approximate the NW cross-section.}
\label{fig:sketch}
\end{figure}

Figure~\ref{fig:sketch} schematically shows a sectorial annular cross-section (SAC). We assume that the NW is oriented parallel to the $z$~axis, thus the cross-section lies in the $x$-$y$~plane. As sketched in Fig.~\ref{fig:sketch}, the four parameters defining the SAC are the inner radius~$R_i \geq 0$, the outer radius~$R_o > R_i$, and the angles $\phi_S$ and $\phi_E$ (with $ 0 < \phi_E - \phi_S < 2 \pi $) at which the cross-section starts and ends, respectively. These four parameters also define the confining potential $V$ [Eq.~(\ref{eq:VforNWConfinement})] for the electrons in our model, because we consider a hard wall at the boundary of the cross-section. The total sectorial angle is given by 
\begin{equation}
\phi_{\rm tot} = \phi_E - \phi_S .
\end{equation}
For all results presented in the main text we choose $\phi_E = \pi - \phi_S$, i.e., the NW cross-section is mirror-symmetric with respect to the $y$ axis. The angles $\phi_S$ and $\phi_E$ are thus related to $\phi_{\rm tot}$ through $\phi_S = (\pi - \phi_{\rm tot})/2$ and $\phi_E = (\pi + \phi_{\rm tot})/2$, respectively.  

We would like to point out that the aforementioned confining potential
\begin{equation}
V = \left\{ \begin{array}{ll}
0 , & R_i < \rho < R_o \mbox{ and } \phi_S < \phi < \phi_E , \\ 
\infty , & \mbox{otherwise} ,
\end{array} \right. 
\label{eq:VforNWConfinement}
\end{equation} 
with $\rho = \sqrt{x^2 + y^2}$, $x = \rho\cos\phi$, and $y = \rho\sin\phi$, is crucial for our quantitative results. By considering a constant potential inside the NW and a hard wall at the boundary, we follow earlier works such as Refs.~\cite{csontos:prb09, nowak:prb13}. However, as briefly explained below, it is also important to note that our assumption is not justified for all devices. For example, given~$V$ of Eq.~(\ref{eq:VforNWConfinement}), a ground-state electron in our model will have a high probability density near the center of the NW cross-section. This holds true if we use $V + \Delta E_{\rm CB}$ instead of $V$, provided that throughout the wire, $|\Delta E_{\rm CB}|$ is much smaller than the ground-state energy of the confined electron. The introduced function $\Delta E_{\rm CB}$ accounts for position-dependent changes of the conduction band edge with respect to its average value inside the NW (defined here as zero). The situation is different when $|\Delta E_{\rm CB}|$ becomes relatively large. For instance, if the conduction band edge decreases near the NW boundary in such a way that the energy of a confined electron in the ground state is lower than the conduction band edge at the center, the electron will have a high probability density near the boundary instead of the center. In such a case, our potential~$V$ should be replaced accordingly. For possible options, see, e.g., the models of Refs.~\cite{kammermeier:prb16, bringer:prb19}. Whether the electrons are mainly localized near the center of the NW or elsewhere can depend on details of the device \cite{jespersen:prb15, heedt:nanoscale15, degtyarev:scirep17}. We believe that the simple approximations made here by using $V$ of Eq.~(\ref{eq:VforNWConfinement}) will be sufficiently justified for many novel devices, particularly when NW cross-sections turn out to be small enough for the ground-state energy to exceed $|\Delta E_{\rm CB}|$ and large enough to avoid significant leakage of the ground-state wave function into the surroundings of the NW. Some suggestions aimed at improving the accuracy of our calculations are described in Sec.~\ref{sec:Discussion}. Finally, we wish to emphasize that many qualitative results in this paper (such as the form of the effective Hamiltonians in Table~\ref{tab:EffDSOI}) do not depend on the specific choice for $V$ and can therefore also be used when Eq.~(\ref{eq:VforNWConfinement}) is not applicable to certain fabricated devices.

\subsection{Hamiltonian without spin-orbit interaction}
\label{secsub:HamWithoutSOI} 

The NWs studied in this work consist of semiconductors with a zinc-blende lattice. We consider materials such as GaAs, InAs, or InSb, where the conduction band minimum is found at the $\Gamma$ point. Inside the NWs, the low-energy electrons are therefore well described by the effective Hamiltonian~\cite{winkler:book}
\begin{equation}
H_0 = \frac{(-i\hbar\nabla)^2}{2 m_{\rm{eff}}} = - \frac{\hbar^2 \Delta}{2 m_{\rm{eff}}} , 
\label{eq:H0}
\end{equation}
where $m_{\rm{eff}}$ is the effective mass, $\nabla$ is the Nabla operator, and $\Delta = \nabla^2$ is the Laplace operator. We omitted here electric and magnetic fields and SOI (see Secs.~\ref{sec:CalculationEffDSOI} to \ref{sec:EffRashbaSOI}). In cylindrical coordinates $\rho, \phi, z$, the Hamiltonian of Eq.~(\ref{eq:H0}) reads  
\begin{equation}
H_0 = - \frac{\hbar^2}{2 m_{\rm{eff}}} \left( \partial^2_\rho + \frac{1}{\rho}\partial_\rho + \frac{1}{\rho^2}\partial^2_\phi + \partial_z^2 \right) .
\label{eq:H0cyl}
\end{equation}
We note that the function
\begin{equation}
\psi = \left[ c_1 J_\alpha (k_\perp \rho) + c_2 Y_\alpha (k_\perp \rho) \right] \left( c_3 e^{i \alpha \phi} + c_4 e^{- i \alpha \phi} \right) e^{i k_z z} 
\label{eq:psiGeneral}
\end{equation}
satisfies 
\begin{equation}
H_0 \psi = \frac{\hbar^2}{2 m_{\rm{eff}}} \left( k_\perp^2 + k_z^2 \right) \psi ,
\label{eq:H0PsiEqualsEtcBulk}
\end{equation}
where $k_\perp$ and $k_z$ are wavenumbers. The $c_i$ in Eq.~(\ref{eq:psiGeneral}) are complex coefficients, the $J_\alpha$ and $Y_\alpha$ stand for Bessel functions of the first and second kind, respectively, and the order of these Bessel functions is denoted by $\alpha$. Remarkably, given the properties of the Bessel functions, Eq.~(\ref{eq:H0PsiEqualsEtcBulk}) is satisfied for an arbitrary complex number $\alpha$.     

Equation~(\ref{eq:psiGeneral}) is of the form 
\begin{equation}
\psi(\rho, \phi, z) = \psi_\perp(\rho, \phi) e^{i k_z z} . 
\end{equation}
The factor $e^{i k_z z}$ is consistent with the translational invariance along the $z$ axis of our model, i.e., with the assumption of an infinitely long NW. Thus, in order to find the low-energy eigenstates of the Hamiltonian $H_0 + V$, we need to choose $\psi_\perp(\rho, \phi)$ such that the hard-wall boundary conditions given by $V$ are fulfilled. In the following, we distinguish between the cases $R_i > 0$ and $R_i = 0$.

\subsection{Nonzero inner radius}
\label{secsub:NonzeroInnerRadius} 

When $R_i > 0$, the boundary conditions 
\begin{equation}
\psi_{\perp}(\rho,\phi_S) = \psi_{\perp}(\rho,\phi_E) = \psi_{\perp}(R_i,\phi) = \psi_{\perp}(R_o,\phi) = 0 
\label{eq:BoundaryCondNonzeroRi}
\end{equation}
must be satisfied for $\rho \in [R_i , R_o]$ and $\phi \in [\phi_S , \phi_E]$. A suitable choice of the coefficients $c_i$ yields
\begin{equation}
\psi_{\perp} = \mathscr{N}_{\perp} \left[ J_\alpha (k_\perp \rho) + C Y_\alpha (k_\perp \rho) \right] \sin\left(\frac{m \pi \phi}{\phi_{\rm tot}} + \varphi_{0,m} \right),
\label{eq:psiperpAnsatzSAC}
\end{equation}
where $m \in \{ 1, 2, 3, \ldots \}$ is a positive integer and 
\begin{eqnarray}
\varphi_{0,m} &=& -\frac{m\pi\phi_S}{\phi_{\rm tot}} , \\
\alpha &=& \frac{m \pi}{\phi_{\rm tot}} . 
\label{eq:alphaFormula}
\end{eqnarray}
The normalization factor $\mathscr{N}_{\perp}$ ensures that
\begin{equation}
\int_{R_i}^{R_o} d\rho \hspace{0.05cm} \rho \int_{\phi_S}^{\phi_E} d\phi \left|\psi_{\perp}\right|^2 = 1 . 
\label{eq:NormalizationCondMainText}
\end{equation}  
Given the boundary conditions, the wavenumber $k_\perp > 0$ and the coefficient $C$ are chosen such that $\psi_{\perp}$ vanishes at $\rho = R_i$ and $\rho = R_o$. For this, the determinant equation 
\begin{equation}
J_\alpha (k_\perp R_i) Y_\alpha (k_\perp R_o) - J_\alpha (k_\perp R_o) Y_\alpha (k_\perp R_i) = 0 
\label{eq:DeterminantEquatForkperp}
\end{equation} 
must be solved. Having found a suitable $k_\perp$, the respective value of $C$ can be calculated. We note that in the limit $|C| \to \infty$, our ansatz [Eq.~(\ref{eq:psiperpAnsatzSAC})] corresponds to a function $\psi_\perp$ whose $\rho$-dependent part contains solely $Y_\alpha (k_\perp \rho)$. However, this special case was not needed for the results presented in this paper. Furthermore, we note that values which differ from those described above, such as negative $k_\perp$ or negative $m$, do not lead to additional (i.e., independent) functions $\psi_\perp$ that are normalizable and satisfy the boundary conditions. 

It is worth mentioning that $J_\alpha (k_\perp \rho)$ and $Y_\alpha (k_\perp \rho)$ are real-valued for real $\alpha$ and $k_\perp \rho > 0$. Consequently, the coefficient $C$ is always real in our calculations, whereas the normalization factor $\mathscr{N}_{\perp}$ is only defined up to an arbitrary phase factor. By choosing $\mathscr{N}_{\perp}$ as real-valued, the function $\psi_{\perp}$ given in Eq.~(\ref{eq:psiperpAnsatzSAC}) is real for $\phi \in [\phi_S , \phi_E]$, $\rho \in [R_i , R_o]$, $k_\perp > 0$ and real $\alpha$ [Eq.~(\ref{eq:alphaFormula})]. In our calculations, however, we never chose a specific phase factor for $\mathscr{N}_{\perp}$, since knowledge of $|\mathscr{N}_{\perp}|^2$ was sufficient for the results presented here.

\subsection{No inner radius}
\label{secsub:NoInnerRadius} 

When $R_i = 0$, $\psi_\perp$ must vanish at $\rho = 0$. However, since $Y_\alpha (k_\perp \rho)$ diverges for $\rho \to 0$, one can set $C = 0$ in Eq.~(\ref{eq:psiperpAnsatzSAC}). Suitable values for $k_\perp$ are therefore simply obtained from $J_\alpha (k_\perp R_o) = 0$ instead of Eq.~(\ref{eq:DeterminantEquatForkperp}). We note that Bessel functions of the first kind have the properties $J_{\alpha=0}(0) = 1$ and $J_{\alpha>0}(0) = 0$. Consequently, as required by the boundary conditions and the continuity of the wave function, $J_\alpha (k_\perp R_i) = J_\alpha (0)$ is always zero because of $\alpha > 0$, see Eq.~(\ref{eq:alphaFormula}). Apart from these small and useful changes for the special case of $R_i = 0$, the wave functions $\psi(\rho, \phi, z) = \psi_\perp(\rho, \phi) e^{i k_z z}$ are calculated exactly as described in Sec.~\ref{secsub:NonzeroInnerRadius} for $R_i > 0$.

\subsection{Eigenenergies and examples}
\label{secsub:EigenenergiesAndExamples} 

At $k_z = 0$, the energy of an electron in the NW is
\begin{equation}
E_{\perp} = \frac{\hbar^2 k_{\perp}^2}{2m_{\rm{eff}}} . 
\label{eq:EperpMainText}
\end{equation}
Thus, having found the eigenstates of $H_0 + V$ at $k_z = 0$, we can order these eigenstates according to their eigenenergies~$E_{\perp}$. The energy gaps between them correspond to the gaps between the subbands of the NW. Since the electron spin is not affected by the Hamiltonian $H_0 + V$, the spin degeneracy can be lifted via additional terms only.     

\begin{figure}[tb]
\centering 
\includegraphics[width=1.00\linewidth]{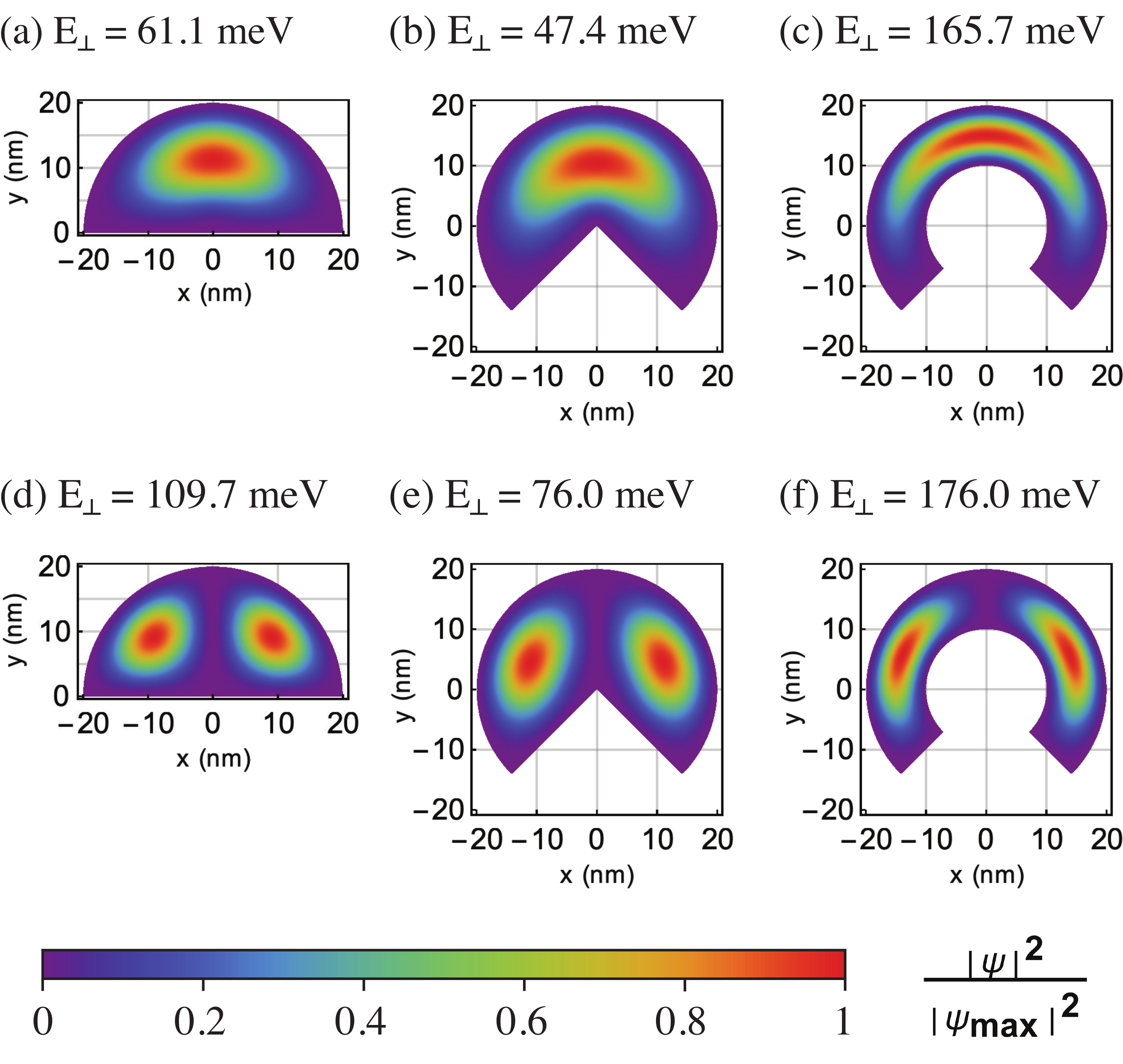} 
\caption{Probability density $|\psi|^2 = |\psi_{\perp}|^2$ for the ground state (top) and the first excited state (bottom) of the Hamiltonian $H_0 + V$ for $k_z = 0$ and three different NW cross-sections. These cross-sections correspond to examples A (panels a and d), B (b and e), C (c and f) and are described in Sec.~\ref{secsub:EigenenergiesAndExamples}. The associated eigenenergies $E_\perp$ were calculated with the effective electron mass of InAs and yield energy gaps of 48.6~meV, 28.6~meV, and 10.3~meV for examples A, B, and C, respectively. For each of the three cross-sections, the ground state was obtained with the number $m = 1$ and the first excited state with $m = 2$. We verified in each case that other eigenstates, particularly excited ones with $m = 1$, have a higher $E_\perp$ than the two states depicted here. The plots show the position dependence of $|\psi|^2/|\psi_{\rm max}|^2$, where $|\psi_{\rm max}|$ is the maximum value of $|\psi|$.     }
\label{fig:SACexample}
\end{figure}

Figure~\ref{fig:SACexample} shows the orbital ground state (top row) and first excited state (bottom row) which we calculated with the Hamiltonian $H_0 + V$ for three different NW cross-sections. More precisely, the probability densities $|\psi|^2 = |\psi_{\perp}|^2$ are plotted for the mentioned states. The three cross-sections in Fig.~\ref{fig:SACexample} have the outer radius $R_o = 20\mbox{ nm}$ and are referred to as examples~A, B, and~C. Example~A corresponds to a half-disk and is obtained by setting $R_i = 0$ and $\phi_{\rm tot} = \pi$. As it will become apparent in Secs.~\ref{sec:CalculationEffDSOI} and \ref{sec:NumericalResults}, a half-disk is a particularly promising NW cross-section for realizing strong DSOI due to its $x$-$y$ confinement ratio. Example~B is defined by $R_i = 0$ and $\phi_{\rm tot} = \frac{3 \pi}{2}$, which is a circular sector of central angle $270^\circ$. Example~C corresponds to a SAC of nonzero inner radius. Its parameter values are $\phi_{\rm tot} = \frac{3 \pi}{2}$ and $R_i = 10\mbox{ nm}$. The eigenenergies and energy gaps provided in Fig.~\ref{fig:SACexample} were calculated with $m_{\rm eff} = 0.0229 m_0$ for InAs \cite{winkler:book}, where $m_0$ is the free electron mass.

\begin{figure}[tb]
\centering 
\includegraphics[width=0.75\linewidth]{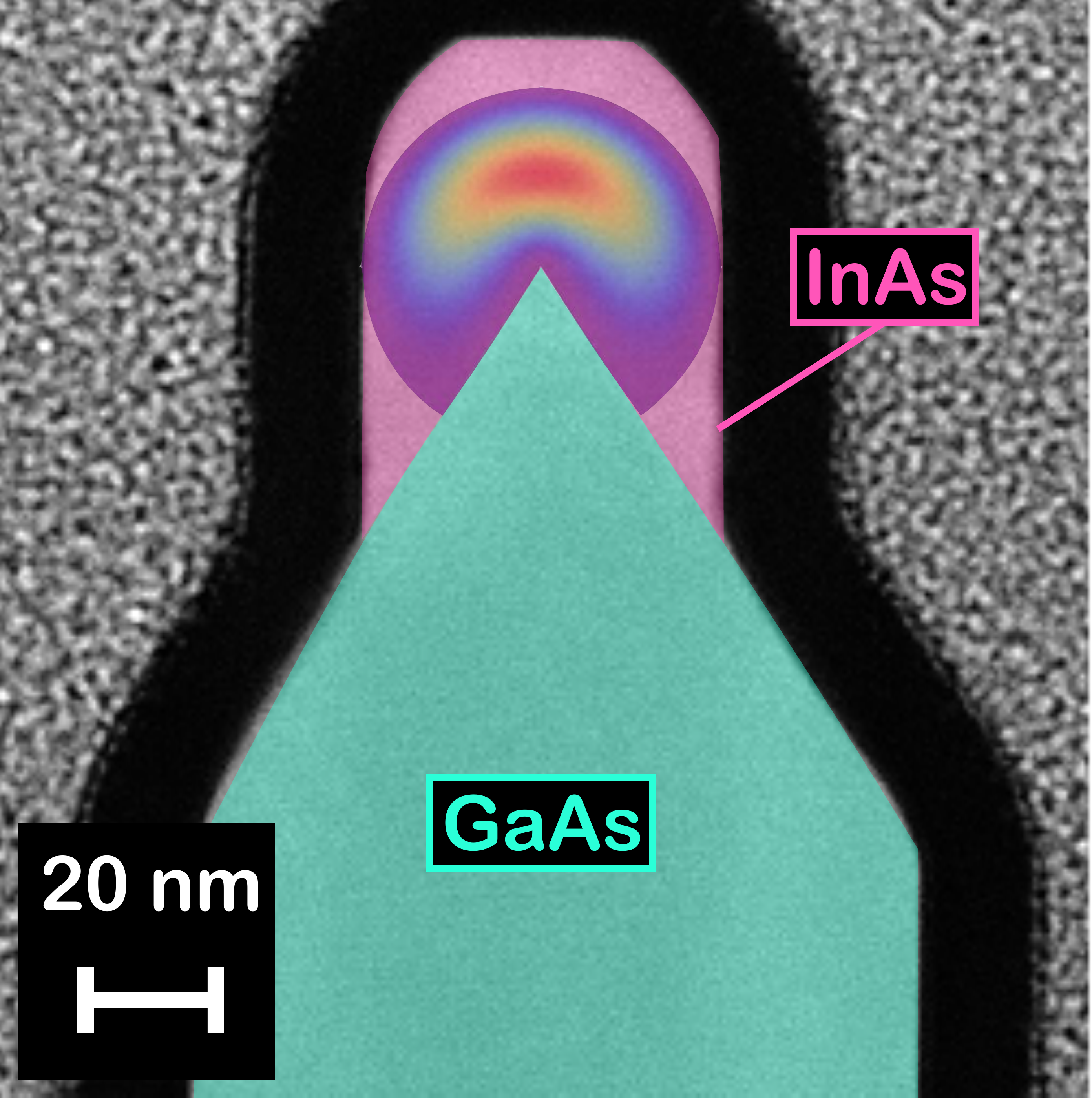} 
\caption{Annular dark field scanning transmission electron microscopy (ADF-STEM) image of a nanomembrane-NW structure grown by Friedl \textit{et al.} \cite{friedl:nl18}. The ADF-STEM image was provided by authors of Ref.~\cite{friedl:nl18} and shows a cross-section of the structure. The InAs NW, whose cross-section we colored here in magenta, formed on top of a GaAs nanomembrane (green). The superimposed image illustrates the probability density (see also Fig.~\ref{fig:SACexample}) of a ground-state electron when the NW cross-section is approximated with a circular sector of about 20~nm radius and $297^\circ$ central angle. }
\label{fig:CrossSecFriedlEtAl}
\end{figure}

The InAs NWs fabricated by Friedl \textit{et al.}~\cite{friedl:nl18} were an important motivation for the present work. These NWs are located on top of GaAs nanomembranes, which were grown on GaAs(111)B substrates. The nanomembranes and NWs are parallel to crystallographic directions of type $\langle 11\bar{2} \rangle$. Based on the results in Ref.~\cite{friedl:nl18}, we now consider a NW along the $[11\bar{2}]$ direction and assume that this NW sits on the $(3\bar{1}1)$ and $(\bar{1}31)$ facets of a nanomembrane. The total sectorial angle in our model is therefore $\phi_{\rm tot} = 2 \pi - 2 \arccos(2\sqrt{2/11}) \approx 1.65 \pi$, which is equivalent to $297^\circ$. Figure~\ref{fig:CrossSecFriedlEtAl} shows the cross-section of a GaAs-InAs nanomembrane-NW structure grown by Friedl \textit{et al.}; it is superimposed by the calculated ground state (analogous to Fig.~\ref{fig:SACexample}) for the parameter values $R_i = 0$, $R_o = 20\mbox{ nm}$, and $\phi_{\rm tot} = 1.65 \pi$.

\section{Calculation of effective Dresselhaus spin-orbit interaction}
\label{sec:CalculationEffDSOI}

\subsection{Orientation of crystallographic axes}
\label{secsub:OrientationCrystalAxes}

\begin{figure}[tb]
\centering
\includegraphics[width=0.70\linewidth]{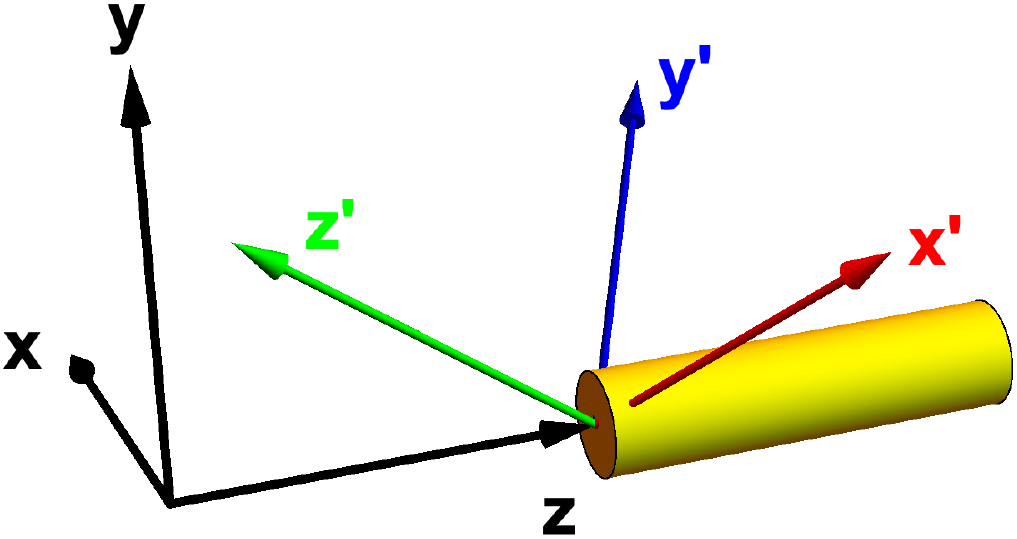} 
\caption{Sketch of the ``noncoincident'' configuration, where \mbox{$x \parallel [\bar{1}10]$}, $y \parallel [111]$, and $z \parallel [11\bar{2}]$. The arrows in this image correspond to normalized vectors and illustrate the orientations of the associated axes (see labels). The axes $x'$, $y'$, and $z'$ coincide with the main crystallographic directions $[100]$, $[010]$, and $[001]$, respectively. The $z$~axis is parallel to the NW. }
\label{fig:wireaxes}
\end{figure}

In order to provide an insight into how the orientation of the crystallographic axes impacts the magnitude of the DSOI, we performed detailed calculations for two different sets of crystallographic basis vectors. In the ``noncoincident'' configuration, the $z$ axis (parallel to the NW) corresponds to the $[11\bar{2}]$ direction, while the $x$ and $y$ axes (see Fig.~\ref{fig:sketch}) correspond to $[\bar{1}10]$ and $[111]$, respectively. This orientation of the crystallographic axes is sketched in Fig.~\ref{fig:wireaxes} and agrees with the NWs of Ref.~\cite{friedl:nl18}. The second configuration, referred to as the ``coincident'' configuration, is obtained when $x$, $y$, and $z$ correspond to the directions $[100]$, $[010]$, and $[001]$, i.e., when the coordinate axes coincide with the main crystallographic axes. We comment on additional configurations in Appendix~\ref{app:DerivationEffDresselhaus}.

\subsection{Effective Dresselhaus term}
\label{secsub:EffDresselhausTerm}

For the semiconductors considered in this work, the DSOI of low-energy electrons in bulk material is \cite{winkler:book, hanson:rmp07} 
\begin{equation}
H_D = b_D \left( \left\{ k_{x'}, k^2_{y'} - k^2_{z'} \right\} \sigma_{x'} + \mbox{c.p.} \right) ,
\label{eq:HDSOI}
\end{equation}
where $x'$, $y'$, and $z'$ are the main crystallographic axes given by the zinc-blende lattice, $\sigma_\nu$ are the Pauli operators for the electron spin, $b_D$ \cite{footnote:bD} is a material-dependent coefficient, $\{A, B\} = (AB + BA)/2$, and the abbreviation ``c.p.''\ stands for cyclic permutations. We keep the notation in this paper simple by using the notation $k_\nu$ both for momentum operators (might also be written as $\hat{k}_\nu$, for example) and wavenumbers (i.e., scalars). 

In order to study the dominant effects of the DSOI in systems with quantum confinement, it is often convenient to derive an effective DSOI term from Eq.~(\ref{eq:HDSOI}), as explained in Ref.~\cite{hanson:rmp07}. For instance, in the special case of a quantum well with strong confinement along the $z'$ axis one obtains an effective Dresselhaus term $H_{D,\rm{eff}} \propto ( k_{y'} \sigma_{y'} - k_{x'} \sigma_{x'} )$ for the low-energy electrons in the quantum well \cite{hanson:rmp07}. Effective Dresselhaus terms for NWs can be derived analogously \cite{kammermeier:prb16, campos:prb18}, see Appendix~\ref{app:DerivationEffDresselhaus} for details and Table~\ref{tab:EffDSOI} for several examples. In summary, we simplify $H_D$ by projecting it onto the two (two because of the spin degree of freedom) lowest-energy subbands of the NW. For this, we compute the average of $H_D$ with respect to the orbital ground-state wave function $\psi_{\perp,g}$ in the $x$-$y$ plane (NW cross-section). This average will be referred to by the short-hand notation 
\begin{equation}
\langle O \rangle = \bra{\psi_{\perp,g}} O \ket{\psi_{\perp,g}}, 
\label{eq:DefinitionLangleRangle} 
\end{equation} 
where $O$ stands for an arbitrary operator. The additional subscript ``$g$'' in $\psi_{\perp,g}$ simply indicates the ground state, i.e., we use here the function $\psi_\perp$ (see Sec.~\ref{sec:NWCrossSectionsBasisStates}) whose associated energy $E_{\perp}$ given in Eq.~(\ref{eq:EperpMainText}) is minimal. We note that an average with respect to $\psi_{\perp,g}$ neither affects the spin operators nor the momentum along the NW. In fact, as briefly mentioned above, $\langle O \rangle$ corresponds to a projection of $O$ onto the two lowest-energy subbands. In the derivation of the effective DSOI terms, we furthermore use $k_\nu k_\mu = k_\mu k_\nu$, meaning that we omit orbital corrections from magnetic fields, if present. Finally, the operator $k_z$ is replaced by the wavenumber $k_z$ (in agreement with the translational invariance along the $z$~axis) and terms proportional to $k_z^3$ are omitted because these are small in the considered regime where $k_z^2 \ll \langle k_{x,y}^2\rangle$. Nevertheless, the $k_z$-cubic terms can be found in Appendix~\ref{app:DerivationEffDresselhaus}, if needed. 

\begin{table}[tb]
\caption{Effective DSOI for different orientations of the crystallographic axes. The $z$~axis is parallel to the NW. The axes $x$ and $y$ are chosen such that $\langle k_x k_y \rangle = 0$. Corrections proportional to $k_z^3$ are listed in Table~\ref{tab:CubicTermsEffDSOI}. Details are provided in Appendix~\ref{app:DerivationEffDresselhaus}. } 
\label{tab:EffDSOI}
\begin{tabular}{|c|c|c|c|}
\hline
$x$ & $y$ & $z$ & $H_{D, {\rm eff}}$ \\
\hline 
$[100]$ & $[010]$ & $[001]$ & $b_D \left( \langle k_x^2 \rangle - \langle k_y^2 \rangle \right) \sigma_z k_z $ \\
$[110]$ & $[\bar{1}10]$ & $[001]$ & $0$ \\
$[001]$ & $[1\bar{1}0]$ & $[110]$ & $ - \frac{b_D}{2} \left( 2 \langle k_x^2 \rangle + \langle k_y^2 \rangle \right) \sigma_y k_z $ \\
$[\bar{1}10]$ & $[001]$ & $[110]$ & $\frac{b_D}{2} \left( \langle k_x^2 \rangle + 2 \langle k_y^2 \rangle \right) \sigma_x k_z$ \\ 
$[\bar{1}11]$ & $[1\bar{1}2]$ & $[110]$ & $\frac{b_D}{2 \sqrt{3}} \left[ \langle k_y^2 \rangle  \left( \sigma_y + 3 \sqrt{2} \sigma_x \right) - 4 \langle k_x^2 \rangle \sigma_y \right] k_z$ \\ 
$[11\bar{2}]$ & $[\bar{1}10]$ & $[111]$ & $\frac{b_D}{\sqrt{6}} \left( \langle k_y^2 \rangle - \langle k_x^2 \rangle \right) \sigma_y k_z$ \\
$[1\bar{1}0]$ & $[11\bar{2}]$ & $[111]$ & $\frac{b_D}{\sqrt{6}} \left( \langle k_y^2 \rangle - \langle k_x^2 \rangle \right) \sigma_x k_z$ \\
$[\bar{1}10]$ & $[111]$ & $[11\bar{2}]$ & $\frac{b_D}{2 \sqrt{3}} \left( \langle k_x^2 \rangle - 4 \langle k_y^2 \rangle \right) \sigma_x k_z $ \\ 
$[111]$ & $[1\bar{1}0]$ & $[11\bar{2}]$ & $\frac{b_D}{2 \sqrt{3}} \left( 4 \langle k_x^2 \rangle - \langle k_y^2 \rangle \right) \sigma_y k_z $ \\
\hline
\end{tabular}
\end{table}

By proceeding as described above, we obtain the effective DSOI term
\begin{equation}
H_{D, {\rm eff}}^{[11\bar{2}]} = \frac{b_D}{2 \sqrt{3}} \left( \langle k_x^2 \rangle - 4 \langle k_y^2 \rangle \right) \sigma_x k_z  = \beta_{\rm eff}  \sigma_x k_z 
\label{eq:HDeff11m2MainText}
\end{equation}
for the noncoincident configuration. The details of the derivation are explained in Appendix~\ref{appsub:DSOIeffNWaxis11m2}. The coefficient $\beta_{\rm eff}$ introduced in Eq.~(\ref{eq:HDeff11m2MainText}) is an effective Dresselhaus parameter (EDP). It solely depends on the NW cross-section and the material-dependent coefficient $b_D$. It can be seen that $\beta_{\rm{eff}}$ vanishes for $\langle k_x^2 \rangle = 4 \langle k_y^2 \rangle$, which can be fulfilled with a cross-sectional confinement that is stronger in the $x$ than in the $y$ direction. As evident from $H_{D, {\rm eff}}^{[11\bar{2}]} = \beta_{\rm eff}  \sigma_x k_z$, the DSOI gives rise to an effective magnetic field parallel to the $x$~axis (see Fig.~\ref{fig:sketch}). For the NWs of Ref.~\cite{friedl:nl18}, this corresponds to an effective magnetic field which is parallel to the substrate (i.e., in-plane) and perpendicular to the NW. The conclusions we can draw from the form of Eq.~(\ref{eq:HDeff11m2MainText}) apply also to recently realized $\langle 11\bar{2} \rangle$-oriented NWs on InP(111)B substrates \cite{aseev:nl19}, for example. In stark contrast to Eq.~(\ref{eq:HDeff11m2MainText}) for the noncoincident configuration, we obtain
\begin{equation}
H_{D, {\rm eff}}^{[001]} = b_D \left( \langle k_x^2 \rangle - \langle k_y^2 \rangle \right) \sigma_z k_z = \bar{\beta}_{\rm eff} \sigma_z k_z 
\label{eq:HDeff001MainText} 
\end{equation}
for the coincident configuration. Here the DSOI leads to an effective magnetic field parallel to the NW. Moreover, the EDP $\bar{\beta}_{\rm eff}$ becomes zero for $\langle k_x^2 \rangle = \langle k_y^2 \rangle$, i.e., for $x:y$ confinement ratios of $1:1$. This is consistent with previous calculations for $\langle 1 0 0 \rangle$-oriented NWs \cite{luo:prb11, kammermeier:prb16, campos:prb18}. Additional information about the effective Dresselhaus term in the case of $z \parallel [001]$ is provided in Appendix~\ref{appsub:DSOIeffNWaxis001}.

\subsection{Scaling properties}
\label{secsub:ScalingProperties}

The EDPs $\beta_{\rm eff}$ and $\bar{\beta}_{\rm eff}$ introduced in Eqs.~(\ref{eq:HDeff11m2MainText}) and (\ref{eq:HDeff001MainText}) have important properties. Given the SAC of Sec.~\ref{secsub:HardWallConf} (Fig.~\ref{fig:sketch}) with $\phi_E = \pi - \phi_S = (\pi + \phi_{\rm tot})/2$, we find
\begin{eqnarray}
\beta_{\rm eff} &=& \frac{b_D}{R_o^2} f(\phi_{\rm tot} , r) , 
\label{eq:betaScalingMainTextNoBar} \\ 
\bar{\beta}_{\rm eff} &=& \frac{b_D}{R_o^2} \bar{f}(\phi_{\rm tot} , r) , 
\label{eq:betaScalingMainTextWithBar} 
\end{eqnarray} 
where the two functions $f$ and $\bar{f}$ depend solely on the total sectorial angle $\phi_{\rm tot} = \phi_E - \phi_S$ and the ratio $r = R_i / R_o$ of inner to outer radius. As expected, Eqs.~(\ref{eq:betaScalingMainTextNoBar}) and (\ref{eq:betaScalingMainTextWithBar}) imply that the EDPs are inversely proportional to the area $R_o^2 (1 - r^2) \phi_{\rm tot} / 2$ of the SAC when $\phi_{\rm tot}$ and $r$ are kept constant. The equations analogously imply that $\beta_{\rm eff} \propto d^{-2}$ and $\bar{\beta}_{\rm eff} \propto d^{-2}$ for any fixed $\phi_{\rm tot}$ and $r$, where
\begin{equation}
d = R_o - R_i = R_o \left( 1 - r \right) 
\end{equation}
is the radial thickness of the SAC. The material dependence of the EDPs results from the proportionality to $b_D$. Due to the hard-wall confinement in our model, the EDPs are independent of the effective mass $m_{\rm eff}$. We note that 
\begin{gather}
\langle k_x^2 \rangle = \frac{4 \bar{f} - 2 \sqrt{3} f}{3 R_o^2} = \frac{4 \bar{\beta}_{\rm eff} - 2 \sqrt{3} \beta_{\rm eff}}{3 b_D} , \\ 
\langle k_y^2 \rangle = \frac{\bar{f} - 2 \sqrt{3} f}{3 R_o^2} = \frac{\bar{\beta}_{\rm eff} - 2 \sqrt{3} \beta_{\rm eff}}{3 b_D} . 
\end{gather} 
Furthermore, we would like to emphasize that $f$ and $\bar{f}$ are dimensionless, which is a convenient property. For details, see Appendix~\ref{app:SizeDependence}.

\subsection{Spin-orbit length and spin-orbit energy}
\label{secsub:SOLengthSOEnergy}

Our model and approximations lead to an effective 1D Hamiltonian of type 
\begin{equation}
H_{\rm 1D} = \frac{\hbar^2 k_z^2}{2 m_{\rm{eff}}} + \beta k_z \sigma_j 
\label{eq:H1DMainText} 
\end{equation}
for the two energetically lowest subbands in the NW. The term $\beta k_z \sigma_j$, where $\beta$ is an EDP and $\sigma_j$ a Pauli operator, corresponds to the effective DSOI, see Sec.~\ref{secsub:EffDresselhausTerm} and Appendix~\ref{app:DerivationEffDresselhaus}. It is well known that the spectrum of the Hamiltonian $H_{\rm 1D}$ in Eq.~(\ref{eq:H1DMainText}) is composed of two parabolas in the energy-$k_z$ diagram \cite{bychkov:jetp84, bychkov:jpcssp84, kloeffel:prb11, kloeffel:prb18}. These parabolas cross at $k_z = 0$ and their minima occur at $k_z = \pm \lambda_{\rm SO}^{-1}$. The spin-orbit length 
\begin{equation}
\lambda_{\rm SO} = \frac{\hbar^2}{m_{\rm{eff}} |\beta|} 
\label{eq:SpinOrbitLength}
\end{equation}
and the spin-orbit energy 
\begin{equation}
E_{\rm SO} = \frac{\hbar^2}{2 m_{\rm{eff}} \lambda_{\rm SO}^2} =  \frac{m_{\rm{eff}} \beta^2}{2 \hbar^2}
\label{eq:SpinOrbitEnergy}
\end{equation}
are two quantities that are of great interest regarding the realization of, among other things, Majorana fermions \cite{alicea:rpp12, beenakker:annurev13, lutchyn:nrm18}, spin filters \cite{streda:prl03}, or quantum logic gates via electric dipole spin resonance \cite{rashba:prl03, golovach:prb06edsr, flindt:prl06, kloeffel:annurev13}. In the next section, we will therefore discuss not only the EDPs but also the spin-orbit lengths and energies obtained with our calculations.

\section{Numerical results}
\label{sec:NumericalResults}

\subsection{Methods and remarks}
\label{secsub:MethodsAndRemarks}

The numerical results presented in Secs.~\ref{secsub:QuantitResNoInnerR} and \ref{secsub:QuantitResNonzeroInnerR} were obtained as follows. For given values of the parameters $R_i$, $R_o$, and $\phi_{\rm tot}$, the function $\psi_\perp$ that belongs to the ground state of the Hamiltonian $H_0 + V$ was calculated as explained in Sec.~\ref{sec:NWCrossSectionsBasisStates}. In order to indicate the ground state, this function is also denoted by $\psi_{\perp,g}$ (Sec.~\ref{secsub:EffDresselhausTerm}). Next, we calculated the expectation values $\langle k_x^2 \rangle = \bra{\psi_{\perp,g}} k_x^2 \ket{\psi_{\perp,g}}$, $\langle k_y^2 \rangle$, and $\langle k_x k_y \rangle$ via numerical integration, using the abovementioned function $\psi_{\perp,g}$ and the operators 
\begin{gather}
k_x = - i \partial_x = - i \cos\phi \hspace{0.06cm} \partial_\rho + i \frac{\sin\phi}{\rho} \partial_\phi , \\
k_y = - i \partial_y = - i \sin\phi \hspace{0.06cm} \partial_\rho - i \frac{\cos\phi}{\rho} \partial_\phi   
\end{gather}
in position-space representation. As a consistency check, we performed the numerical integration both in Cartesian and cylindrical coordinates. Apart from tiny differences due to the finite numerical precision, the results from both methods were always identical. Moreover, $\langle k_x k_y \rangle$ always vanished. We note that $\langle k_x k_y \rangle = 0$ is indeed expected because of the mirror symmetry of the cross-section with respect to the $y$ axis. For a discussion on how strongly $\langle k_x k_y \rangle$ usually depends on the choice of the axes $x$ and $y$, we refer to Appendix~\ref{app:ChoiceOfAxes}. Finally, having evaluated $\langle k_x^2 \rangle$ and $\langle k_y^2 \rangle$ for the given parameter values, we calculated
\begin{gather}
f(\phi_{\rm tot} , r) = \frac{R_o^2}{2 \sqrt{3}} \left( \langle k_x^2 \rangle - 4 \langle k_y^2 \rangle \right) , \\
\bar{f}(\phi_{\rm tot} , r) = R_o^2 \left( \langle k_x^2 \rangle - \langle k_y^2 \rangle \right) .
\end{gather}
In agreement with Appendix~\ref{app:SizeDependence}, we obtained the same values (apart from tiny differences related to the numerical precision) for $f$ or $\bar{f}$, respectively, when $R_i$ and $R_o$ were changed such that their ratio $r = R_i/R_o$ remained constant. The results for $f$ and $\bar{f}$, which are dimensionless and material-independent, also enabled us to calculate the EDPs $\beta_{\rm eff}$ and $\bar{\beta}_{\rm eff}$ [Eqs.~(\ref{eq:betaScalingMainTextNoBar}) and (\ref{eq:betaScalingMainTextWithBar})] and, furthermore, the associated spin-orbit lengths [Eq.~(\ref{eq:SpinOrbitLength})] and energies [Eq.~(\ref{eq:SpinOrbitEnergy})]. For these material-dependent quantities, we considered InAs and chose $m_{\rm eff} = 0.0229 m_0$ and $b_D = 27.18\mbox{ nm$^3$ meV}$ \cite{winkler:book, footnote:bD}. In Sec.~\ref{secsub:ConversionFactors}, we provide conversion factors with which our results for InAs can easily be adapted to other semiconductors such as InSb.           

The expectation values of the operators $k_x$, $k_y$, $k_x^3$, $k_y^3$, $k_x k_y^2$, and $k_x^2 k_y$ must vanish because the electrons are trapped inside the NW. In the derivation of the effective DSOI terms (Appendix~\ref{app:DerivationEffDresselhaus} and Sec.~\ref{secsub:EffDresselhausTerm}), we thus set $\langle k_{x,y} \rangle = \langle k_{x,y}^3 \rangle = \langle k_{x,y}^2 k_{y,x} \rangle = 0$. By evaluating these expectation values numerically as a consistency check, we became aware of artifacts in the results for $\langle k_y^3 \rangle$ and  $\langle k_x^2 k_y \rangle$. As explained in Appendix~\ref{app:ArtifactHardWall}, these artifacts are not caused by the numerical integration; they arise from the hard-wall boundary conditions, which generally allow for wave functions with discontinuous derivatives at the interfaces, and the fact that the considered NW cross-sections have no mirror symmetry with respect to an axis parallel to the $x$~axis. Fortunately, the numerically calculated $\langle k_x^2 \rangle$ and $\langle k_y^2 \rangle$ are free of such artifacts, which justifies our assumption of hard-wall confinement in order to gain insight into the effective DSOI. Furthermore, we note that the evaluated expectation values $\langle k_{x,y} \rangle$, $\langle k_x^3 \rangle$, and $\langle k_x k_y^2 \rangle$ always vanished, as expected. For detailed information, see Appendix~\ref{app:ArtifactHardWall}.

\subsection{No inner radius}
\label{secsub:QuantitResNoInnerR} 

\begin{figure}[tb]
\centering 
\includegraphics[width=0.86\linewidth]{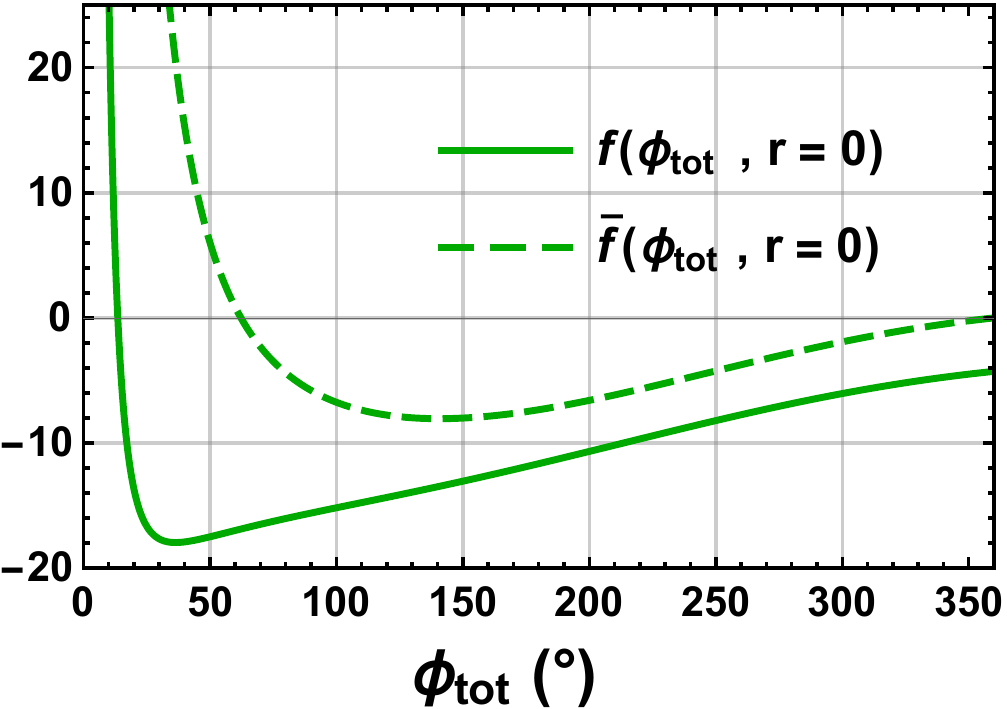} 
\caption{Angular dependence of the functions $f(\phi_{\rm tot} , r)$ (solid line) and $\bar{f}(\phi_{\rm tot} , r)$ (dashed line) at $r = 0$, in which case the NW cross-section is a circular sector of central angle $\phi_{\rm tot}$. For small angles $\phi_{\rm tot} \lesssim 13^\circ$, it turns out that $\langle k_x^2 \rangle > 4 \langle k_y^2 \rangle$ at $r = 0$, and so $f(\phi_{\rm tot} , 0)$ and $\bar{f}(\phi_{\rm tot} , 0)$ are both positive. In contrast, both $f(\phi_{\rm tot} , 0)$ and $\bar{f}(\phi_{\rm tot} , 0)$ are negative at large angles, because with $r = 0$ one finds $\langle k_x^2 \rangle < \langle k_y^2 \rangle$ for $\phi_{\rm tot} \gtrsim 63^\circ$. As expected, $f$ and $\bar{f}$ diverge when $\phi_{\rm tot}$ vanishes, since $\langle k_x^2 \rangle \to \infty$ and $\langle k_y^2 \rangle  / \langle k_x^2 \rangle \to 0$ for $\phi_{\rm tot} \to 0$.}
\label{fig:fbarfForZeroInnerRadius}
\end{figure}

If $R_i = r = 0$, the cross-section of the NW is a circular sector with central angle $\phi_{\rm tot}$. The functions $f(\phi_{\rm tot} , 0)$ and $\bar{f}(\phi_{\rm tot} , 0)$ for this special case are plotted in Fig.~\ref{fig:fbarfForZeroInnerRadius}. In combination with the equations provided in Sec.~\ref{sec:CalculationEffDSOI} and Appendix~\ref{app:DerivationEffDresselhaus}, the data in Fig.~\ref{fig:fbarfForZeroInnerRadius} allows to quickly obtain an estimate of the effective DSOI for any radius $R_o$, any central angle $\phi_{\rm tot}$, and any of the discussed growth directions (see, e.g., Table~\ref{tab:EffDSOI}).         

Some examples with $R_i = r = 0$ are listed in Table~\ref{tab:ResultsNumericalCalc}, where we focus on three different values for $\phi_{\rm tot}$. First, $\phi_{\rm tot} = 1.65\pi = 297^\circ$ is of particular interest because this angle applies to a NW that forms on the $(3\bar{1}1)$ and $(\bar{1}31)$ facets of a nanomembrane \cite{friedl:nl18}, see Sec.~\ref{secsub:EigenenergiesAndExamples} and Fig.~\ref{fig:CrossSecFriedlEtAl}. The second value $\phi_{\rm tot} = \frac{3 \pi}{2} = 270^\circ$ may be used as a relatively simple approximation for various structures. For instance, a $[11\bar{2}]$-oriented NW might alternatively be grown on a nanomembrane with $(5\bar{1}2)$ and $(\bar{1}52)$ facets, leading to a central angle of $\phi_{\rm tot} = 1.564\pi = 281.5^\circ$, or a $[001]$-oriented NW might in principle be grown on $(\bar{1}10)$ and $(110)$ facets, in which case a central angle of exactly $270^\circ$ would be expected. In fact, it turns out that our EDPs for $\phi_{\rm tot} = 297^\circ$ and $\phi_{\rm tot} = 270^\circ$ differ by less than a factor of two, so the latter angle can also serve as a reasonable approximation for the NWs of Ref.~\cite{friedl:nl18}. The third value $\phi_{\rm tot} = \pi = 180^\circ$ leads to a cross-section that corresponds to a half-disk. As evident from Table~\ref{tab:ResultsNumericalCalc}, a relatively strong DSOI is obtained for this NW shape.

\begin{table*}[tb]
\centering

\caption{Numerical results for $f$ ($\bar{f}$), the EDP $\beta_{\rm{eff}}$ ($\bar{\beta}_{\rm{eff}}$), and the associated spin-orbit length $\lambda_{\rm{SO}}$ ($\bar{\lambda}_{\rm{SO}}$) and energy $E_{\rm{SO}}$ ($\bar{E}_{\rm{SO}}$) obtained in the noncoincident (coincident) configuration. The listed examples correspond to NW cross-sections with a radial thickness $d$ of 4~nm, 10~nm, or 20~nm. We find that $f$, $\bar{f}$, $\beta_{\rm{eff}}$, and $\bar{\beta}_{\rm{eff}}$ are negative for all examples in this table. It is important to note that the EDPs, the spin-orbit lengths, and the spin-orbit energies are material-dependent. The results listed here were calculated for InAs as explained in the text. They can easily be adapted to other semiconductors via the conversion factors in Sec.~\ref{secsub:ConversionFactors}. }
\label{tab:ResultsNumericalCalc}

\begin{tabular}{|c|c|c|c|c||c|c||c|c|c||c|c|c|}
\hline
\mbox{ $\phi_{\rm{tot}}$ } & $r$ & $R_i$ & $R_o$ & $d$ & $|f|$ & $|\bar{f}|$ &
$|\beta_{\rm{eff}}|$ & $\lambda_{\rm{SO}}$ & $E_{\rm{SO}}$ &
$|\bar{\beta}_{\rm{eff}}|$ & $\bar{\lambda}_{\rm{SO}}$ & $\bar{E}_{\rm{SO}}$ \\
 
\mbox{  } &  & \mbox{ (nm) } & \mbox{ (nm) } & \mbox{ (nm) } & & & \mbox{ (meV nm) } & \mbox{ ($\mu$m) } & \mbox{ ($\mu$eV) } & \mbox{ (meV nm) } & \mbox{ ($\mu$m) } & \mbox{ ($\mu$eV) }
 
\\ \hline \hline
 
\mbox{ 180$^\circ$ } & 0 & 0 & 4 & 4 & 11.7 & 7.4 & 19.8 & 0.17 & 59.0 & 12.5 & 0.27 & 23.4 \\ \hline

270$^\circ$ & 0 & 0 & 4 & 4 & 7.3 & 3.3 & 12.4 & 0.27 & 23.1 & 5.5 & 0.60 & 4.62 \\ \hline

297$^\circ$ & 0 & 0 & 4 & 4 & 6.2 & 2.0 & 10.5 & 0.32 & 16.5 & 3.5 & 0.96 & 1.80 \\ \hline

270$^\circ$ & 0.2 & 1 & 5 & 4 & 9.4 & 3.4 & 10.2 & 0.33 & 15.7 & 3.7 & 0.91 & 2.02 \\ \hline

270$^\circ$ & 0.6 & 6 & 10 & 4 & 34.7 & 10.8 & 9.4 & 0.35 & 13.4 & 2.9 & 1.13 & 1.30 \\ \hline

270$^\circ$ & 0.8 & 16 & 20 & 4 & \mbox{ 137.3 } & \mbox{ 42.1 } & 9.3 & 0.36 & 13.1 & 2.9 & 1.16 & 1.23 \\ \hline

180$^\circ$ & 0 & 0 & 10 & 10 & 11.7 & 7.4 & 3.2 & 1.05 & 1.51 & 2.0 & 1.67 & 0.60 \\ \hline

270$^\circ$ & 0 & 0 & 10 & 10 & 7.3 & 3.3 & 2.0 & 1.68 & 0.59 & 0.89 & 3.75 & 0.12 \\ \hline

297$^\circ$ & 0 & 0 & 10 & 10 & 6.2 & 2.0 & 1.7 & 1.98 & 0.42 & 0.55 & 6.05 & 0.05 \\ \hline

270$^\circ$ & \mbox{ 0.091 } & 1 & 11 & 10 & 7.8 & 3.1 & 1.8 & 1.90 & 0.46 & 0.69 & 4.81 & 0.07 \\ \hline

270$^\circ$ & 0.5 & 10 & 20 & 10 & 22.4 & 7.1 & 1.5 & 2.19 & 0.35 & 0.48 & 6.88 & 0.04 \\ \hline

180$^\circ$ & 0 & 0 & 20 & 20 & 11.7 & 7.4 & 0.79 & 4.20 & 0.09 & 0.50 & 6.67 & 0.04 \\ \hline

270$^\circ$ & 0 & 0 & 20 & 20 & 7.3 & 3.3 & 0.50 & 6.71 & 0.04 & 0.22 & $> 10$ & 0.01 \\ \hline

297$^\circ$ & 0 & 0 & 20 & 20 & 6.2 & 2.0 & 0.42 & 7.92 & 0.03 & 0.14 & $> 10$ & $< 0.01$ \\ \hline

\end{tabular}
\end{table*}

\subsection{Nonzero inner radius}
\label{secsub:QuantitResNonzeroInnerR} 

\begin{figure}[b]
\centering 
\includegraphics[width=0.86\linewidth]{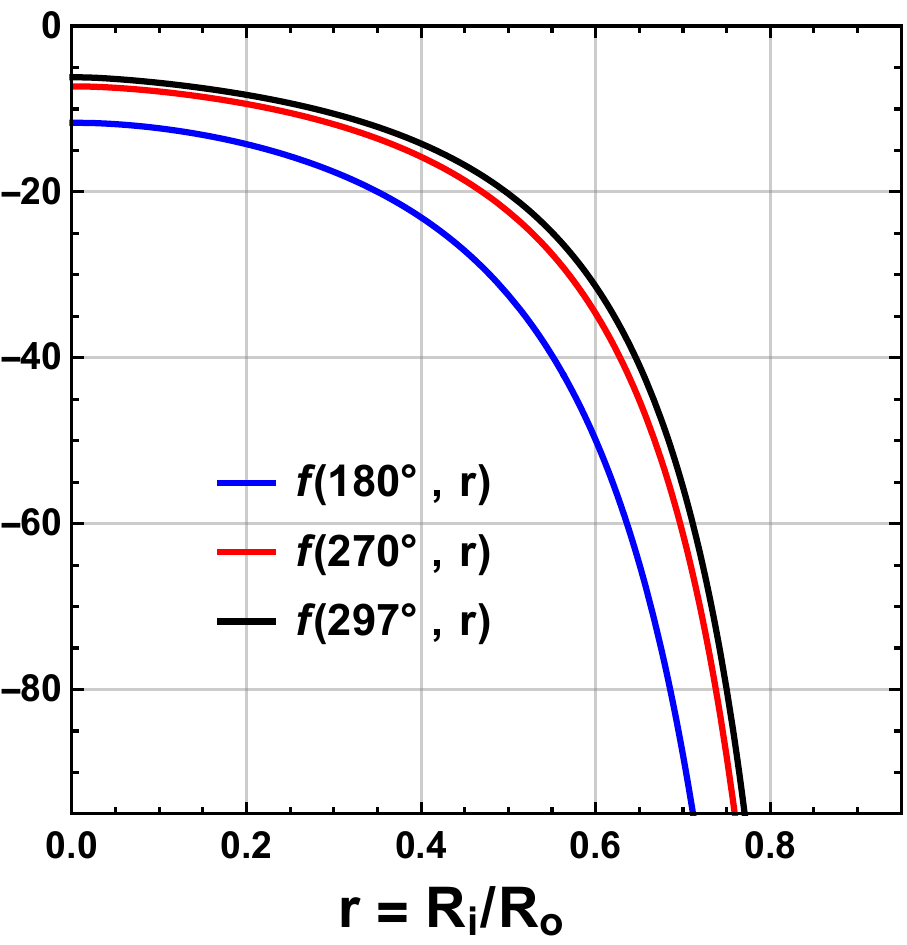} 
\caption{Dependence of the function $f(\phi_{\rm tot} , r)$ on the ratio $r$ of inner to outer radius for the values $180^\circ$ (blue), $270^\circ$ (red), and $297^\circ$ (black) of the angle $\phi_{\rm tot}$. The results for $r \to 0$ are consistent with the solid line in Fig.~\ref{fig:fbarfForZeroInnerRadius}. It turns out that $f$ and $\bar{f}$ (Fig.~\ref{fig:fbarForThreeAngles}) diverge when \mbox{$r \to 1$}, in agreement with the expected behavior of $\langle k_x^2 \rangle$ and $\langle k_y^2 \rangle$ when \mbox{$d/R_o \to 0$}. }
\label{fig:fForThreeAngles}
\end{figure}

\begin{figure}[b]
\centering 
\includegraphics[width=0.86\linewidth]{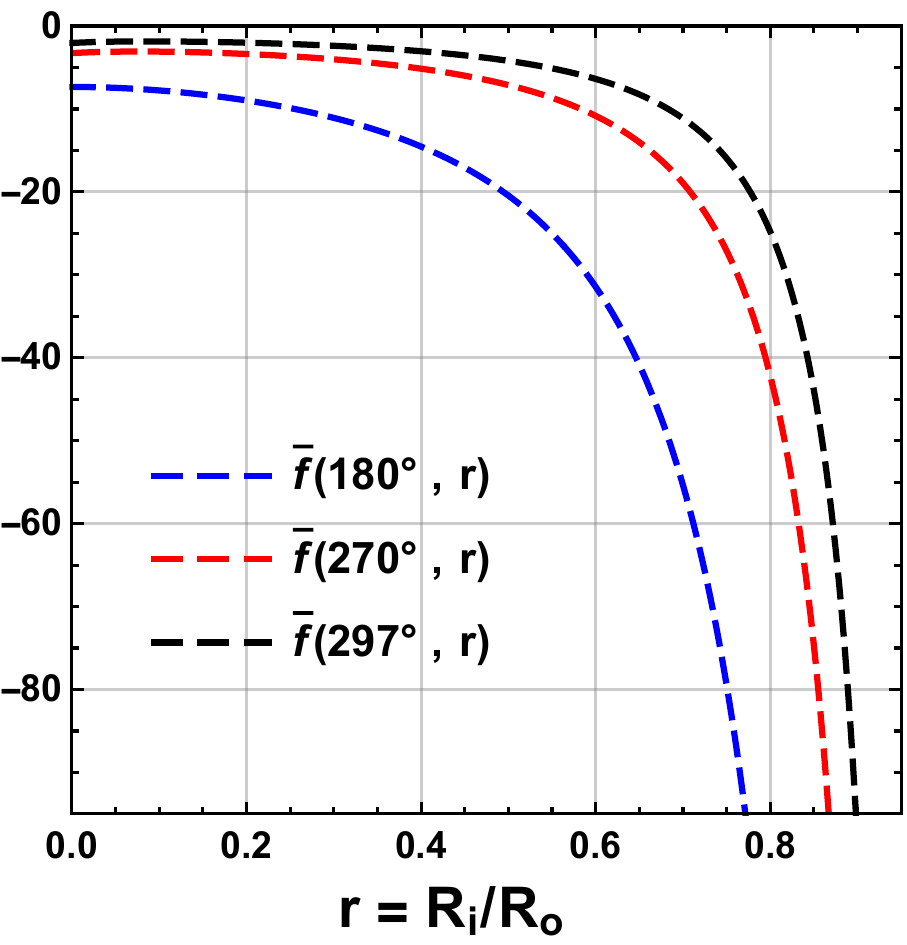} 
\caption{Dependence of $\bar{f}(\phi_{\rm tot} , r)$ on $r$ for the values $180^\circ$ (blue), $270^\circ$ (red), and $297^\circ$ (black) of $\phi_{\rm tot}$. The results shown here correspond to the coincident configuration and may be directly compared with the related results for $f(\phi_{\rm tot} , r)$ in Fig.~\ref{fig:fForThreeAngles}, corresponding to the noncoincident configuration. For $r \to 0$, the values of $\bar{f}(\phi_{\rm tot} , r)$ plotted here are consistent with the dashed line in Fig.~\ref{fig:fbarfForZeroInnerRadius}.}
\label{fig:fbarForThreeAngles}
\end{figure}

For the three special values of $\phi_{\rm tot}$ discussed in Sec.~\ref{secsub:QuantitResNoInnerR}, the calculated data in Figs.~\ref{fig:fForThreeAngles} and \ref{fig:fbarForThreeAngles} show the dependence of $f(\phi_{\rm tot} , r)$ and $\bar{f}(\phi_{\rm tot} , r)$, respectively, on the ratio $r$. Examples for associated spin-orbit lengths and energies in the case of InAs NWs are listed in Table~\ref{tab:ResultsNumericalCalc}. We note that the continuum model (envelope function approximation, $\bm{k} \cdot \bm{p}$ theory \cite{winkler:book}) employed in Sec.~\ref{sec:NWCrossSectionsBasisStates} will eventually lose validity if the area of the NW cross-section is reduced until it is based on a few atoms only. We therefore set $d = R_o - R_i \geq 4\mbox{ nm}$ in all examples presented here. 

It is evident from the numbers in Table~\ref{tab:ResultsNumericalCalc} that small NW cross-sections are needed in order to obtain a strong SOI which originates from the DSOI in InAs. This holds true for both the noncoincident and the coincident configuration. The main reason for this result is the fact that the Dresselhaus coefficient $b_D$ of InAs is not extraordinarily large, even though InAs has a rather narrow energy gap between the lowest conduction band and the highest valence band. For instance, the values of $b_D$ obtained perturbatively from an extended Kane model for the semiconductors GaAs, AlAs, InAs, CdTe, and ZnSe are all in the range 10--50~nm$^3$~meV \cite{winkler:book, footnote:bD}. As a consequence, it may not be surprising that the authors of Ref.~\cite{friedl:nl18} concluded from their magnetotransport measurements that the SOI was weak, with an estimated lower bound of 280~nm for the spin-orbit length. The authors also mentioned that a stronger SOI may be achieved in future devices by using InSb NWs. In Sec.~\ref{secsub:ConversionFactors}, we therefore provide conversion factors. Compared with the abovementioned semiconductors, InSb has a remarkably large Dresselhaus coefficient $b_D$ of about 760~nm$^3$~meV \cite{winkler:book, footnote:bD}. By analyzing our results (e.g., Table~\ref{tab:ResultsNumericalCalc}) also for InSb, we conclude that an unusually strong DSOI, with associated spin-orbit energies above 1~meV, is possible with InSb NWs for both the noncoincident and the coincident configuration.

\subsection{Conversion factors for other semiconductors}
\label{secsub:ConversionFactors} 

Results for a material $X$ can immediately be adapted to a material $Y$ via the relations 
\begin{gather}
\beta_{\rm eff}^Y = \beta_{\rm eff}^X \frac{b_D^Y}{b_D^X}  = \beta_{\rm eff}^X  \kappa^{X \to Y}_{\beta} , \label{eq:ConversionNonCoincBeta} \\ 
\lambda_{\rm SO}^Y = \lambda_{\rm SO}^X \frac{m_{\rm eff}^X b_D^X}{m_{\rm eff}^Y b_D^Y} = \lambda_{\rm SO}^X  \kappa^{X \to Y}_{\lambda} , \label{eq:ConversionNonCoincLambdaSO} \\ 
E_{\rm SO}^Y = E_{\rm SO}^X \frac{m_{\rm eff}^Y \left( b_D^Y \right)^2}{m_{\rm eff}^X \left( b_D^X \right)^2}  = E_{\rm SO}^X  \kappa^{X \to Y}_{E} , \label{eq:ConversionNonCoincESO}
\end{gather}
where the superscript added to $\beta_{\rm eff}$, $\lambda_{\rm SO}$, $E_{\rm SO}$, $b_D$, and $m_{\rm eff}$ indicates the material. The three dimensionless factors $\kappa^{X \to Y}_{\beta, \lambda, E}$ are conversion factors for the EDP, the spin-orbit length, and the spin-orbit energy, respectively. We note that the equations $\bar{\beta}_{\rm eff}^Y = \bar{\beta}_{\rm eff}^X  \kappa^{X \to Y}_{\beta}$, $\bar{\lambda}_{\rm SO}^Y = \bar{\lambda}_{\rm SO}^X \kappa^{X \to Y}_{\lambda}$, and $\bar{E}_{\rm SO}^Y = \bar{E}_{\rm SO}^X  \kappa^{X \to Y}_{E}$ for the coincident configuration are identical to Eqs.~(\ref{eq:ConversionNonCoincBeta}), (\ref{eq:ConversionNonCoincLambdaSO}), and (\ref{eq:ConversionNonCoincESO}) for the noncoincident configuration. 

As evident from Eqs.~(\ref{eq:ConversionNonCoincBeta}) to (\ref{eq:ConversionNonCoincESO}), the introduced conversion factors depend on the effective masses and the Dresselhaus coefficients of the materials. The effective mass $m_{\rm eff}$ of a semiconductor is usually well known. In contrast, reported values for the Dresselhaus coefficient $b_D$~\cite{footnote:bD} often vary quite strongly. Throughout this paper, we use the material parameters listed in Ref.~\cite{winkler:book}. We note, however, that our results can easily be recalculated with other values, if desired. For example, several methods have been developed with which $b_D$ can be extracted from experimental data \cite{jusserand:prb95, knap:prb96, miller:prl03, krich:prl07, faniel:prb11, walser:nphys12, walser:prb12, dettwiler:prx17, weigele:arX18, stano:prb18, stano:prb19, marinescu:prl19}. 

With the parameters \cite{winkler:book, footnote:bD} $b_D^{\rm InSb} = 760.1\mbox{ nm$^3$ meV}$, $b_D^{\rm InAs} = 27.18\mbox{ nm$^3$ meV}$, $m_{\rm eff}^{\rm InSb} = 0.0139 m_0$, and $m_{\rm eff}^{\rm InAs} = 0.0229 m_0$, one obtains the conversion factors 
\begin{eqnarray}
\kappa^{\rm InAs \to InSb}_{\beta} &=& 27.97 , \label{eq:ConvFactBetaInAsToInSb} \\
\kappa^{\rm InAs \to InSb}_{\lambda} &=& 0.0589 , \\
\kappa^{\rm InAs \to InSb}_{E} &=& 474.7 .
\end{eqnarray}
By replacing InAs with InSb, we thus find that the EDPs $\beta_{\rm eff}$ and $\bar{\beta}_{\rm eff}$ in Table~\ref{tab:ResultsNumericalCalc} increase by a factor of about thirty, that the spin-orbit lengths $\lambda_{\rm SO}$ and $\bar{\lambda}_{\rm SO}$ shorten by a factor of about twenty, and that the spin-orbit energies $E_{\rm SO}$ and $\bar{E}_{\rm SO}$ increase by two to three orders of magnitude. In stark contrast, using \cite{winkler:book, footnote:bD} $b_D^{\rm GaAs} = 27.58\mbox{ nm$^3$ meV}$ and $m_{\rm eff}^{\rm GaAs} = 0.0665 m_0$ yields \begin{eqnarray}
\kappa^{\rm InAs \to GaAs}_{\beta} &=& 1.015 , \\
\kappa^{\rm InAs \to GaAs}_{\lambda} &=& 0.339 , \\
\kappa^{\rm InAs \to GaAs}_{E} &=& 2.99 ,
\end{eqnarray}        
so replacing InAs with GaAs would have rather small effects on the results in Table~\ref{tab:ResultsNumericalCalc}. As a consequence, according to the material parameters in Ref.~\cite{winkler:book}, only minor quantitative differences are expected between identically shaped InAs, GaAs, and InGaAs NWs regarding the DSOI. Large differences between these NWs, on the other hand, are expected regarding the RSOI (see Sec.~\ref{sec:EffRashbaSOI}). Finally, we would like to mention that by changing from InAs to InSb or GaAs, the energies given in Fig.~\ref{fig:SACexample} are rescaled by a factor of $m_{\rm eff}^{\rm InAs}/m_{\rm eff}^{\rm InSb} = 1.65$ or $m_{\rm eff}^{\rm InAs}/m_{\rm eff}^{\rm GaAs} = 0.344$, respectively.  

It is important to note that we have thus far focused on the Dresselhaus contribution to the SOI. The Rashba term will be considered next.

\section{Effective Rashba spin-orbit interaction}
\label{sec:EffRashbaSOI} 

The RSOI of electrons is described by a term of type 
\begin{equation}
H_R = a_R \left( \bm{k} \times \bm{E} \right) \cdot \bm{\sigma} = a_R \left( \bm{\sigma} \times \bm{k} \right) \cdot \bm{E} ,  
\end{equation}
where $a_R$ is a Rashba coefficient \cite{footnote:aR}, $\bm{\sigma}$ is the vector of Pauli matrices, and $\bm{E}$ is an effective electric field that accounts for the structure inversion asymmetry of the confining potential \cite{bychkov:jetp84, bychkov:jpcssp84, winkler:book}. In stark contrast to DSOI, the RSOI Hamiltonian does not depend on the orientation of the crystallographic axes. Projecting $H_R$ onto the two lowest subbands of the NW yields the effective RSOI term    
\begin{equation}
H_{R, {\rm eff}} = \langle H_R \rangle = a_R \left( E_x \sigma_y - E_y \sigma_x \right) k_z  
\end{equation} 
for the low-energy electrons. In combination with the effective DSOI terms derived in Sec.~\ref{secsub:EffDresselhausTerm}, we thus obtain
\begin{equation}
H_{D, {\rm eff}}^{[11\bar{2}]} + H_{R, {\rm eff}} = \left[ \left( \beta_{\rm eff} - a_R E_y \right) \sigma_x + a_R E_x \sigma_y \right] k_z  
\label{eq:SumHDHREffFor11m2}   
\end{equation}
for the noncoincident configuration and
\begin{equation}
H_{D, {\rm eff}}^{[001]} + H_{R, {\rm eff}} = \left(- a_R E_y \sigma_x + a_R E_x \sigma_y + \bar{\beta}_{\rm eff} \sigma_z \right) k_z 
\label{eq:SumHDHREffFor001}  
\end{equation}
for the coincident configuration. The components $E_x$ and $E_y$ of the effective electric field $\bm{E}$ inside the NW can be controlled via electric gates in the experimental setup. 

Equation~(\ref{eq:SumHDHREffFor11m2}) describes the SOI of low-energy electrons in the recently grown NWs of Ref.~\cite{friedl:nl18}. As briefly explained below, the predicted SOI in these NWs may be very useful for applications. If the cross-section (i.e., the associated confining potential) of the NW is mirror-symmetric with respect to the $y$~axis, as sketched in Fig.~\ref{fig:sketch}, and if the same applies to the externally induced potential (modifiable via gate voltages), one finds $E_x = 0$. Consequently, Eq.~(\ref{eq:SumHDHREffFor11m2}) simplifies to $( \beta_{\rm eff} - a_R E_y ) \sigma_x k_z$, which corresponds to an electrically tunable SOI proportional to $\sigma_x$. Moreover, since the Dresselhaus and Rashba contributions have the same form, the effective SOI can in principle be set to zero even if $\beta_{\rm eff}$ is nonzero (DSOI and RSOI cancel each other). By tuning $E_x$ and/or $E_y$ via electric gates, the SOI may then be changed from zero to a desired form considering Eq.~(\ref{eq:SumHDHREffFor11m2}). In the coincident configuration, for example, the effective SOI cannot be set to zero unless $E_x$, $E_y$, and $\bar{\beta}_{\rm eff}$ all vanish, as evident from Eq.~(\ref{eq:SumHDHREffFor001}).    

Our results for $\beta_{\rm eff}$ and $\bar{\beta}_{\rm eff}$ in Sec.~\ref{sec:NumericalResults} reveal that EDPs of about $10\mbox{ meV nm} = 10^{-11}\mbox{ eVm}$ are possible with InAs and GaAs NWs. Using the value $a_R^{\rm InAs} = 1.2 \mbox{ nm}^2 e$ \cite{winkler:book, footnote:aR} for InAs, we note that $a_R^{\rm InAs} \left| \bm{E} \right| = 10^{-11}\mbox{ eVm}$ is satisfied with $\left| \bm{E} \right| = 8.5\mbox{ V/$\mu$m}$, i.e., with a moderate electric field. In stark contrast to the Dresselhaus coefficients $b_D^{\rm GaAs}$ and $b_D^{\rm InAs}$, which are almost equivalent (see Sec.~\ref{secsub:ConversionFactors} and Ref.~\cite{knap:prb96}), the Rashba coefficients $a_R^{\rm GaAs} = 0.052\mbox{ nm}^2 e$ and $a_R^{\rm InAs}$ differ by a factor of about twenty \cite{winkler:book}. Consequently, a stronger electric field $\left| \bm{E} \right| = 0.19\mbox{ V/nm}$ is needed in order to achieve $a_R^{\rm GaAs} \left| \bm{E} \right| = 10^{-11}\mbox{ eVm}$ for GaAs. These fields below $1\mbox{ V/nm}$ are feasible with electric gates located near the NWs. 

In the case of InSb NWs, we can make use of Eq.~(\ref{eq:ConvFactBetaInAsToInSb}), so our results in Sec.~\ref{sec:NumericalResults} suggest that EDPs of about $3\times 10^{-10}\mbox{ eVm}$ are possible. This example corresponds to a remarkably high spin-orbit energy of about 8~meV due to DSOI, despite the small effective mass $m_{\rm eff}^{\rm InSb} = 0.0139 m_0$. By setting $a_R^{\rm InSb} = 5.2 \mbox{ nm}^2 e$ \cite{winkler:book, footnote:aR}, one finds $a_R^{\rm InSb} \left| \bm{E} \right| = 3\times 10^{-10}\mbox{ eVm}$ at $\left| \bm{E} \right| = 0.06 \mbox{ V/nm}$, which is feasible. For comparison, $a_R^{\rm InSb} \left| \bm{E} \right| = 10^{-11}\mbox{ eVm}$ is satisfied at $\left| \bm{E} \right| = 1.9 \mbox{ V/$\mu$m}$ already. Since the EDPs decrease rapidly when the size of the NW cross-section is increased, as explained in Sec.~\ref{secsub:ScalingProperties} and Appendix~\ref{app:SizeDependence}, it turns out that even for InSb (large $b_D$) NWs of medium-sized cross-section, electric fields of the order of V/$\mu$m are usually sufficient to induce a RSOI which is stronger than the effective DSOI term. Our results adapted to medium-sized cross-sections are thus consistent with the calculations by Campos \textit{et al.}~\cite{campos:prb18}, who studied the RSOI and DSOI in zinc-blende InSb NWs which have hexagonal cross-sections and widths of several tens of nanometers. The authors pointed out that the RSOI clearly dominates in these NWs when an electric field of 4~V/$\mu$m is applied. For small cross-sections, such as those with $d \leq 10\mbox{ nm}$ in Table~\ref{tab:ResultsNumericalCalc}, we find that the DSOI can be the main contribution to the effective SOI even in the presence of electric fields of the order of V/$\mu$m.

\section{Discussion}
\label{sec:Discussion} 

One of the main aims of our work for this paper was to gain information about the SOI of electrons in the novel NWs fabricated by Friedl~\textit{et al.}~\cite{friedl:nl18}. By allowing for a nonzero inner radius, analyzing the parameter dependence, and considering different growth directions, the calculations were extended such that our results can be readily adapted to a large variety of NWs. For example, the introduced functions $f$ and $\bar{f}$ are useful because they are dimensionless and material-independent. Therefore, given our results, it is straightforward to recalculate quantities such as the spin-orbit length and energy, if needed, for any desired material parameters, even if these differ from the material parameters of Ref.~\cite{winkler:book} considered here. For instance, values for the Dresselhaus coefficient $b_D$ \cite{footnote:bD} may be chosen based on related experimental data \cite{jusserand:prb95, knap:prb96, miller:prl03, krich:prl07, faniel:prb11, walser:nphys12, walser:prb12, dettwiler:prx17, weigele:arX18, stano:prb18, stano:prb19, marinescu:prl19}.     

For the $\langle 11\bar{2} \rangle$-oriented InAs NWs of Ref.~\cite{friedl:nl18}  (an example is shown in Fig.~\ref{fig:CrossSecFriedlEtAl}), we find that the effective DSOI is weak when the radius is about 20~nm, as evident from the last row in Table~\ref{tab:ResultsNumericalCalc}. However, we also find that the nanomembrane-NW structures of Ref.~\cite{friedl:nl18} allow for a strong DSOI with an associated spin-orbit energy of the order of meV, provided that the NWs can be made of InSb and their cross-sections can be scaled down. Moreover, by applying an electric field in the out-of-plane direction (perpendicular to the substrate), the induced RSOI enables a cancellation of Rashba and Dresselhaus contributions, so the resulting SOI term can be switched on and off. Our estimates show that the electric field needed for this switching would be well below V/nm, even for the strong DSOI mentioned above, and may therefore be applied via electric gates. As explained in Sec.~\ref{sec:EffRashbaSOI}, such a cancellation of RSOI and DSOI would not be possible for the coincident configuration. The NW networks of Refs.~\cite{friedl:nl18, aseev:nl19}, to which the noncoincident configuration in our model applies, are therefore promising platforms for applications which require an electrically controllable SOI. 

In addition to the RSOI discussed in Sec.~\ref{sec:EffRashbaSOI}, an electric field $\bm{E}$ applied perpendicular to a NW leads to a term $e (E_x x + E_y y)$ in the Hamiltonian, where $e$ is the elementary positive charge. As a consequence, the electron is pushed towards the boundary of the NW cross-section, affecting also the expectation values of operators such as $k_x^2$ and $k_y^2$. For weak and moderate electric fields, the ground-state wave functions obtained in the absence and presence of $\bm{E}$ usually do not differ significantly~\cite{devries:nl18}, thus one may assume for simplicity that $\langle k_x^2 \rangle$ and $\langle k_y^2 \rangle$ remain unchanged when electric fields are applied. However, more accurate results will be obtained when the effects of $\bm{E}$ on $\langle k_x^2 \rangle$ and $\langle k_y^2 \rangle$ are fully taken into account. These corrections generally depend not only on $\bm{E}$ but also on the effective electron mass and the details of the NW cross-section. In some cases, accounting for the electric-field-induced changes of $\langle k_x^2 \rangle$ and $\langle k_y^2 \rangle$ may even be crucial, particularly when $\bm{E}$ is relatively strong. Let us consider, for example, a cylindrical NW and the coincident configuration, i.e., $z \parallel [001]$ for the NW axis, $x \parallel [100]$, and $y \parallel [010]$. The cylindrical symmetry of the confining potential leads to $\langle k_x k_y \rangle = 0$ and $\langle k_x^2 \rangle = \langle k_y^2 \rangle$, and so the effective DSOI term $H_{D, {\rm eff}} = b_D ( \langle k_x^2 \rangle - \langle k_y^2 \rangle ) \sigma_z k_z$ (see, e.g., Table~\ref{tab:EffDSOI}) vanishes. By applying an electric field in the $x$ or $y$ direction, $\langle k_x^2 \rangle > \langle k_y^2 \rangle$ or $\langle k_y^2 \rangle > \langle k_x^2 \rangle$ can be achieved, resulting in a nonzero $H_{D, {\rm eff}}$. This means that the Dresselhaus contribution to the effective SOI can be turned on and off via $\bm{E}$. If the electric field is so strong that the electrons are pushed far into one half of the circular cross-section, $\langle k_x^2 \rangle$ and $\langle k_y^2 \rangle$ may be estimated via one of the low-symmetry cross-sections (e.g., a half-disk) considered in this paper. To some extent, our results are therefore also applicable to NWs where the symmetry is broken by strong electric fields. However, a detailed analysis of how exactly the ground states and associated expectation values $\langle k_x^2 \rangle$ and $\langle k_y^2 \rangle$ depend on $\bm{E}$ and the NW properties remains an open task. More suggestions aimed at improving the accuracy are described below. 

Future calculations may address mechanisms and corrections which were beyond the scope of the present paper. For example, we used here the parameter values of bulk semiconductors \cite{winkler:book}. In the presence of strong confinement, adapted values may be chosen in order to obtain more precise results. In general, corrections which originate from other NW subbands and other bands of the semiconductor may be included \cite{winkler:book}. If available, detailed information about the given system may be taken into account, such as the strain distribution, changes in the material composition, and the details of the confining potential (see also Sec.~\ref{secsub:HardWallConf}). Nonuniform strain, for instance, causes position-dependent shifts of the band edges of the semiconductor \cite{birpikus:book, adachi:book1992, adachi:book2005, winkler:book}, leading not only to rescaled band structure parameters but also to a modified confinement potential for the electrons in the NW. Sophisticated numerical methods and tools are probably necessary in order to study all these corrections. Furthermore, it is important to note that we focused here on SOI which originates from bulk and structure inversion asymmetry \cite{winkler:book}. Additional contributions to the SOI can arise from interface inversion asymmetry \cite{winkler:book, ivchenko:prb96, vervoort:prb97, guettler:prb98, vervoort:sst99, olesberg:prb01, hall:prb03, golub:prb04, nestoklon:prb08, prada:njp11, furthmeier:ncomm16, wojcik:apl19}. It would therefore be very interesting to analyze these contributions for various NWs and interfaces and combine them with our results. For purely wurtzite GaAs/AlGaAs core/shell NWs, for instance, interface-induced SOI was found to be of high relevance \cite{furthmeier:ncomm16}. Recent calculations for InAs/InAsP core/shell NWs suggest that interface-related contributions to the SOI will also be important for many zinc-blende NW heterostructures \cite{wojcik:apl19}.

A special result of our work is evident from Table~\ref{tab:EffDSOI}. Provided that the axes $x$ and $y$ are defined such that $\langle k_x k_y \rangle = 0$ is satisfied, it turns out that $H_{D, {\rm eff}} = 0$ for $x \parallel [110]$, $y \parallel [\bar{1}10]$, $z \parallel [001]$. Furthermore, $k_z$-cubic terms are absent in $H_D$ for this configuration (see the appended Table~\ref{tab:CubicTermsEffDSOI}). We wish to emphasize that these results are independent of $\langle k_x^2 \rangle$ and $\langle k_y^2 \rangle$. The DSOI is therefore strongly suppressed even if electric fields are present, provided that they are applied in such a way that $\langle k_x k_y \rangle = 0$ is conserved. The relation $\langle k_x k_y \rangle = 0$ holds true, e.g., if the NW cross-section (more precisely, the associated confining potential) is mirror-symmetric with respect to the $x$ or $y$ axis and $\bm{E}$ is applied parallel to this axis. 

In conclusion, there are two promising strategies when one wants to switch the SOI in a NW on and off by applying an electric field $\bm{E}$ which is controllable via electric gates. In the first case, the setup is chosen such that the $\bm{E}$-independent contributions (see, e.g., the DSOI terms in Secs.~\ref{sec:CalculationEffDSOI} and \ref{sec:NumericalResults}) to the effective SOI are nonzero and can be cancelled out via the $\bm{E}$-induced contributions (see, e.g., the RSOI terms in Sec.~\ref{sec:EffRashbaSOI}). In the second case, the choices are made such that without $\bm{E}$, the effective SOI is suppressed. The resulting SOI is then fully determined by the terms induced by~$\bm{E}$. We note that in the first (second) case, a nonzero~$\bm{E}$ is needed to turn the effective SOI off (on). For both strategies, however, it is essential to understand how the SOI depends on the specifics of the experimental setup. Our results in this paper can contribute to such an understanding, particularly when novel NWs with low-symmetry cross-sections are used.

\begin{acknowledgments} 
We thank Martin Friedl, Kris Cerveny, Pirmin Weigele, Sara Mart\'{i}-S\'{a}nchez, Taras Patlatiuk, Jordi Arbiol, and Anna Fontcuberta i Morral for helpful discussions and for providing us with the ADF-STEM image in Fig.~\ref{fig:CrossSecFriedlEtAl}. We acknowledge financial support from the Swiss National Science Foundation and the NCCR QSIT. 
\end{acknowledgments}

\appendix

\section{Derivation of the effective Dresselhaus term}
\label{app:DerivationEffDresselhaus}

In this appendix, we show the derivation of the effective Dresselhaus term for low-energy electrons in NWs. The derivation is analogous to the case of 2D-like systems, which is explained in detail in Ref.~\cite{hanson:rmp07}. Results will be provided for four important growth directions. Related calculations can be found, for instance, in Refs.~\cite{kammermeier:prb16, campos:prb18, bringer:prb19}. 

We start from Eq.~(\ref{eq:HDSOI}), see Sec.~\ref{secsub:EffDresselhausTerm}, which is the DSOI for low-energy electrons in a bulk semiconductor with zinc-blende structure and conduction-band edge at the $\Gamma$~point, provided that the s-like $\Gamma_6^c$ is the lowest-lying conduction band \cite{winkler:book}. Neglecting corrections from magnetic fields, one can assume that the operators $k_\nu$ for the electron momentum commute, i.e., $k_\nu k_\mu = k_\mu k_\nu$, and so Eq.~(\ref{eq:HDSOI}) simplifies to     
\begin{equation}
H_D = b_D \left[ \sigma_{x'} k_{x'} \left( k^2_{y'} - k^2_{z'} \right)  + \mbox{c.p.} \right] .
\label{eq:HDSOIwithoutBfields}
\end{equation}
By projection of $H_D$ onto the NW subbands of lowest energy, we obtain the effective DSOI term 
\begin{equation}
H_{D, {\rm eff}} = \langle H_D \rangle = \bra{\psi_{\perp,g}} H_D \ket{\psi_{\perp,g}} , 
\label{eq:HDSOIeffNWgeneral}
\end{equation}
where $\psi_{\perp,g}$ is the orbital ground-state wave function in the \mbox{$x$-$y$} plane. As discussed in the following, the result for $H_{D, {\rm eff}}$ depends strongly on the growth direction of the NW. We will refer to the unit vectors along the axes $x$, $y$, $z$ as $\bm{e}_x$, $\bm{e}_y$, $\bm{e}_z$, respectively. The $z$ axis is parallel to the NW. The unit vectors $\bm{e}_{x'}$, $\bm{e}_{y'}$, $\bm{e}_{z'}$ for the axes $x'$, $y'$, $z'$ point in the main crystallographic directions [100], [010], [001]. We consider right-handed systems, so $\bm{e}_x \times \bm{e}_y = \bm{e}_z$ and $\bm{e}_{x'} \times \bm{e}_{y'} = \bm{e}_{z'}$ (analogously for cyclic permutations).

\subsection{NW axis along $[11\bar{2}]$} 
\label{appsub:DSOIeffNWaxis11m2}

When the NW axis coincides with the $[11\bar{2}]$ direction, the unit vectors are related by 
\begin{equation}
\bm{e}_z = \frac{1}{\sqrt{6}} \left( \bm{e}_{x'} + \bm{e}_{y'} - 2 \bm{e}_{z'}  \right).
\end{equation}
Furthermore, we choose $x \parallel [\bar{1}10]$ and $y \parallel [111]$,
\begin{eqnarray}
\bm{e}_x &=& \frac{1}{\sqrt{2}} \left( \bm{e}_{y'} - \bm{e}_{x'}  \right) , \\
\bm{e}_y &=& \frac{1}{\sqrt{3}} \left( \bm{e}_{x'} + \bm{e}_{y'} + \bm{e}_{z'}  \right) . 
\end{eqnarray}
This choice leads to the relations 
\begin{eqnarray}
k_{x'} &=& - \frac{k_x}{\sqrt{2}} + \frac{k_y}{\sqrt{3}} + \frac{k_z}{\sqrt{6}} , \label{eq:kxprimeForz11m2}\\
k_{y'} &=& \frac{k_x}{\sqrt{2}} + \frac{k_y}{\sqrt{3}} + \frac{k_z}{\sqrt{6}} , \\
k_{z'} &=& \frac{k_y}{\sqrt{3}} - \frac{2 k_z}{\sqrt{6}} 
\label{eq:kzprimeForz11m2}
\end{eqnarray}
between the operators for the momentum, which can be verified via the identity $\bm{e}_{x} k_{x} + \bm{e}_{y} k_{y} + \bm{e}_{z} k_{z} = \bm{e}_{x'} k_{x'} + \bm{e}_{y'} k_{y'} + \bm{e}_{z'} k_{z'}$. Equations~(\ref{eq:kxprimeForz11m2}) to (\ref{eq:kzprimeForz11m2}) also apply to the Pauli operators $\sigma_\nu$ for the spin. With the derived relations for $k_\nu$ and $\sigma_\nu$, the Hamiltonian $H_D$ of Eq.~(\ref{eq:HDSOIwithoutBfields}) is rewritten as
\begin{eqnarray}
H_D &=& \frac{b_D \sigma_x}{\sqrt{3}} \Biggl[ 
\frac{k_x^2 k_y}{\sqrt{2}} + \frac{k_x^2 k_z}{2} - 2 k_y^2 k_z - \frac{k_y k_z^2}{\sqrt{2}} + \frac{k_z^3}{2} \Biggr]  \nonumber \\
& & + \frac{b_D \sigma_y}{\sqrt{6}} \Bigl[ 3 k_x k_z^2 - k_x^3  \Bigr]  \nonumber \\
& & + \frac{b_D \sigma_z}{\sqrt{3}} \Biggl[ 2 k_x k_y^2 - \sqrt{2} k_x k_y k_z - \frac{k_x k_z^2}{2} - \frac{k_x^3}{2} \Biggr] . 
\end{eqnarray}
The effective DSOI term, Eq.~(\ref{eq:HDSOIeffNWgeneral}), can now be obtained easily by making the substitutions
\begin{eqnarray} 
k_{x,y}^2 k_z &\to& \langle k_{x,y}^2 \rangle k_z , \\ 
k_x k_y k_z &\to& \langle k_x k_y \rangle k_z , \\ 
k_{x,y}k_z^2 &\to& \langle k_{x,y} \rangle k_z^2 = 0 , \\ 
k_{x,y}^3 &\to& \langle k_{x,y}^3 \rangle = 0 , \\
k_{x,y}^2 k_{y,x} &\to& \langle k_{x,y}^2 k_{y,x} \rangle = 0 .
\end{eqnarray} 
Furthermore, terms proportional to $k_z^3$ are negligible in the regime of small $k_z$ considered here, because these terms are much smaller than those of type $\langle k_{x,y}^2 \rangle k_z$. In conclusion, we keep only the terms which are linear in $k_z$ and find 
\begin{equation}
H_{D, {\rm eff}}^{[11\bar{2}]} = \frac{b_D k_z}{2 \sqrt{3}} \left[ \left( \langle k_x^2 \rangle - 4 \langle k_y^2 \rangle \right) \sigma_x - 2\sqrt{2} \langle k_x k_y \rangle \sigma_z \right] . 
\label{eqApp:HDeff11m2theta0}
\end{equation}
We note that it is usually possible to choose the orthogonal axes $x$ and $y$ for the transverse directions such that $\langle k_x k_y \rangle = 0$. A simple example is discussed in Appendix~\ref{app:ChoiceOfAxes}. In particular, our calculations revealed that $\langle k_x k_y \rangle = 0$ for all systems studied in the main text, see Fig.~\ref{fig:sketch} for a sketch of the NW cross-section and the considered orientation of the axes. By choosing the axes $x$ and $y$ such that $\langle k_x k_y \rangle = 0$, the effective DSOI term has the compact form   
\begin{equation}
H_{D, {\rm eff}}^{[11\bar{2}]} = \frac{b_D}{2 \sqrt{3}} \left( \langle k_x^2 \rangle - 4 \langle k_y^2 \rangle \right) \sigma_x k_z  .
\label{eqApp:HDeff11m2theta0kxky0}
\end{equation}
This equation is used in the main text [Eq.~(\ref{eq:HDeff11m2MainText})] and describes the DSOI in recently fabricated NWs \cite{friedl:nl18, aseev:nl19}. Thus, unless $ \langle k_x^2 \rangle \simeq 4 \langle k_y^2 \rangle $, the Dresselhaus Hamiltonian leads to a notable SOI in these NWs.

By introducing an angle $\theta$, the most general relations between the main crystallographic directions and the axes $x$, $y$, $z$ are given by
\begin{eqnarray}
\bm{e}_x &=& \frac{\cos\theta}{\sqrt{2}} \left(\bm{e}_{y'} - \bm{e}_{x'} \right) + \frac{\sin\theta}{\sqrt{3}} \left(\bm{e}_{x'} + \bm{e}_{y'} + \bm{e}_{z'} \right) ,  \\
\bm{e}_y &=& \frac{\sin\theta}{\sqrt{2}} \left(\bm{e}_{x'} - \bm{e}_{y'} \right) + \frac{\cos\theta}{\sqrt{3}} \left(\bm{e}_{x'} + \bm{e}_{y'} + \bm{e}_{z'} \right)  ,  \\
\bm{e}_z &=& \frac{1}{\sqrt{6}} \left( \bm{e}_{x'} + \bm{e}_{y'} - 2 \bm{e}_{z'}  \right)
\end{eqnarray}
when the NW axis $z$ corresponds to the $[11\bar{2}]$ direction. With these relations, one obtains the effective DSOI term 
\begin{widetext}
\begin{eqnarray}
H_{D, {\rm eff}}^{[11\bar{2}]} &=& \frac{b_D k_z}{8 \sqrt{3}} \left[ \left( 11 \langle k_x^2 \rangle + \langle k_y^2 \rangle \right) \sigma_y \sin\theta - \left( \langle k_x^2 \rangle + 11 \langle k_y^2 \rangle \right) \sigma_x \cos\theta + 10 \langle k_x k_y \rangle \left( \sigma_y \cos\theta - \sigma_x \sin\theta \right) \right] \nonumber \\
& & + \frac{5 b_D k_z}{8 \sqrt{3}} \left[ \left( \langle k_x^2 \rangle - \langle k_y^2 \rangle \right) \left( \sigma_x \cos(3 \theta) - \sigma_y \sin(3 \theta) \right) - 2 \langle k_x k_y \rangle \left( \sigma_x \sin(3\theta) + \sigma_y \cos(3\theta) \right) \right] \nonumber \\ 
& & + \frac{b_D k_z }{\sqrt{6}} \left[ \left( \langle k_y^2 \rangle - \langle k_x^2 \rangle \right) \sin(2 \theta) - 2 \langle k_x k_y \rangle \cos(2\theta) \right] \sigma_z  . 
\label{eqApp:HDeff11m2LongResultWithAngle}
\end{eqnarray} 
As expected, the result in Eq.~(\ref{eqApp:HDeff11m2theta0}) for the special case of $x \parallel [\bar{1}10]$ ($y \parallel [111]$) is retrieved by setting $\theta = 0$. For example, the angle $\theta = \pi/2$ corresponds here to $x \parallel [111]$ ($y \parallel [1\bar{1}0]$). In the derivation of Eq.~(\ref{eqApp:HDeff11m2LongResultWithAngle}), the $k_z$-cubic term $b_D k_z^3 (\sigma_x \cos\theta - \sigma_y \sin\theta)/(2 \sqrt{3})$ was omitted. 
\end{widetext}

\subsection{NW axis along $[001]$} 
\label{appsub:DSOIeffNWaxis001}

We now consider a setup with $z \parallel [001]$, so $\bm{e}_x = \bm{e}_{x'} \cos\theta + \bm{e}_{y'} \sin\theta$, $\bm{e}_y = \bm{e}_{y'} \cos\theta - \bm{e}_{x'} \sin\theta$, and $\bm{e}_z = \bm{e}_{z'}$. Consequently, $k_{x'} = k_x \cos\theta - k_y \sin\theta$, $k_{y'} = k_y \cos\theta + k_x \sin\theta$, $k_{z'} = k_z$, and the identical relations apply to $\sigma_\nu$. Proceeding analogously to Appendix~\ref{appsub:DSOIeffNWaxis11m2} yields the effective DSOI term 
\begin{eqnarray}
H_{D, {\rm eff}}^{[001]} &=& b_D \left( \langle k_x^2 \rangle - \langle k_y^2 \rangle \right) \cos(2 \theta) \sigma_z k_z  \nonumber \\ 
& & - 2 b_D \langle k_x k_y \rangle \sin(2 \theta) \sigma_z k_z  .
\end{eqnarray}
It is worth noting that $H_D$ [Eq.~(\ref{eq:HDSOIwithoutBfields})] does not contain any terms proportional to $k_z^3$ if $z \parallel [001]$. The special case $x \parallel [100]$ ($y \parallel [010]$), where $x$, $y$, $z$ coincide with main crystallographic directions, is obtained at $\theta = 0$, leading to the simple expression for $H_{D, {\rm eff}}^{[001]}$ shown in Eq.~(\ref{eq:HDeff001MainText}).

\subsection{NW axis along $[110]$} 
\label{appsub:DSOIeffNWaxis110}

In the main text, we focus on NWs oriented along $[11\bar{2}]$ or $[001]$. In this appendix, we consider [110]-oriented NWs for comparison. The relations between the unit vectors are now of the form 
\begin{eqnarray}
\bm{e}_x &=& \left( \bm{e}_{x'} - \bm{e}_{y'} \right) \frac{\sin\theta}{\sqrt{2}} + \bm{e}_{z'} \cos\theta ,  \\
\bm{e}_y &=& \left( \bm{e}_{x'} - \bm{e}_{y'} \right) \frac{\cos\theta}{\sqrt{2}} - \bm{e}_{z'} \sin\theta ,  \\
\bm{e}_z &=& \frac{1}{\sqrt{2}} \left( \bm{e}_{x'} + \bm{e}_{y'} \right)
\end{eqnarray}
and lead to the effective DSOI term
\begin{eqnarray}
H_{D, {\rm eff}}^{[110]} &=& \frac{3 b_D k_z}{8} \Bigl( \langle k_x^2 \rangle - \langle k_y^2 \rangle \Bigr) \Bigl( \sigma_x \sin(3 \theta) + \sigma_y \cos(3 \theta) \Bigr)  \nonumber \\ 
& & + \frac{3 b_D k_z}{4} \langle k_x k_y \rangle \Bigl( \sigma_x \cos(3 \theta) - \sigma_y \sin(3 \theta) \Bigr)   \nonumber \\
& & + \frac{5 b_D k_z}{4} \langle k_x k_y \rangle \Bigl( \sigma_x \cos\theta + \sigma_y \sin\theta \Bigr)  \nonumber \\
& & - \frac{b_D k_z}{8}  \left( \langle k_x^2 \rangle + 11 \langle k_y^2 \rangle \right) \sigma_x \sin\theta  \nonumber \\
& & - \frac{b_D k_z}{8} \left( 11 \langle k_x^2 \rangle + \langle k_y^2 \rangle \right) \sigma_y \cos\theta  .  
\label{eqApp:HDeff110LongResultWithAngle}
\end{eqnarray}
For example, at $\theta = 0$ the result simplifies to
\begin{equation}
H_{D, {\rm eff}}^{[110]} =  \frac{b_D k_z}{2} \left[ 4 \langle k_x k_y \rangle \sigma_x - \left( 2 \langle k_x^2 \rangle + \langle k_y^2 \rangle \right) \sigma_y \right] ,   
\end{equation}
which applies to the case where $x \parallel [001]$ and $y \parallel [1\bar{1}0]$. Setting $\theta = - \arccos(1/\sqrt{3}) = - \arctan(\sqrt{2})$ leads to a configuration where $x \parallel [\bar{1}11]$ and $y \parallel [1\bar{1}2]$. In the derivation of Eq.~(\ref{eqApp:HDeff110LongResultWithAngle}), we omitted the term $b_D k_z^3 (\sigma_y \cos\theta + \sigma_x \sin\theta)/2$ contained in $H_D$.

\subsection{NW axis along $[111]$} 
\label{appsub:DSOIeffNWaxis111}

The fourth case considered in this appendix is described by the relations 
\begin{eqnarray}
\bm{e}_x &=& \frac{\cos\theta}{\sqrt{6}} \left( \bm{e}_{x'} + \bm{e}_{y'} - 2 \bm{e}_{z'}  \right) + \frac{\sin\theta}{\sqrt{2}} \left( \bm{e}_{y'} - \bm{e}_{x'}  \right) , \\
\bm{e}_y &=& \frac{\sin\theta}{\sqrt{6}} \left( 2 \bm{e}_{z'} - \bm{e}_{x'} - \bm{e}_{y'} \right) + \frac{\cos\theta}{\sqrt{2}} \left( \bm{e}_{y'} - \bm{e}_{x'}  \right) , \\
\bm{e}_z &=& \frac{1}{\sqrt{3}} \left( \bm{e}_{x'} + \bm{e}_{y'} + \bm{e}_{z'}  \right)  
\end{eqnarray}
for a NW with $z \parallel [111]$. We proceed again analogously to Appendix~\ref{appsub:DSOIeffNWaxis11m2} and obtain 
\begin{eqnarray}
H_{D, {\rm eff}}^{[111]} &=& \frac{b_D k_z}{\sqrt{6}} \Bigl( \langle k_y^2 \rangle - \langle k_x^2 \rangle \Bigr) \Bigl( \sigma_x \sin(3 \theta) + \sigma_y \cos(3 \theta) \Bigr)  \nonumber \\ 
& & + \frac{2 b_D k_z}{\sqrt{6}}  \langle k_x k_y \rangle \Bigl(\sigma_y \sin(3 \theta) - \sigma_x \cos(3 \theta) \Bigr)  .  
\label{eqApp:HDeff111LongResultWithAngle}
\end{eqnarray}
Like in the case of $z \parallel [001]$ studied in Appendix~\ref{appsub:DSOIeffNWaxis001}, it turns out that there are no $k_z$-cubic terms in $H_D$, Eq.~(\ref{eq:HDSOIwithoutBfields}), if $z \parallel [111]$. The right-hand side of Eq.~(\ref{eqApp:HDeff111LongResultWithAngle}) is a relatively short expression given that the NW axis does not coincide with a main crystallographic direction. As expected from symmetry considerations, the result is invariant when the angle $\theta$ is changed by multiples of $2\pi/3$. Setting $\theta = 0$ corresponds here to $x \parallel [11\bar{2}]$ and $y \parallel [\bar{1}10]$.

\subsection{Summary and remarks} 
\label{appsub:DSOIeffSummary} 

The effective DSOI terms for $\langle k_x k_y \rangle = 0$ (see also Appendix~\ref{app:ChoiceOfAxes}) and commonly used growth directions are listed in Table~\ref{tab:EffDSOI}. If $\langle k_x^2 \rangle  = \langle k_y^2 \rangle$, which is satisfied for some highly symmetric NW cross-sections (e.g., circles or squares), the effective DSOI term $H_{D, {\rm eff}}$ is nonzero for NWs oriented along $[110]$ or $[11\bar{2}]$ but vanishes for NWs oriented along $[001]$ or $[111]$, which is consistent with previous calculations \cite{luo:prb11, kammermeier:prb16, campos:prb18}. We note that recent calculations for electrons confined close to the surface of a cylindrical NW showed that DSOI is relevant for $[111]$-oriented NWs under certain conditions \cite{bringer:prb19}. 

An eye-catching item in Table~\ref{tab:EffDSOI} is the simple result \mbox{$H_{D, {\rm eff}} = 0$} at $x \parallel [110]$, $y \parallel [\bar{1}10]$, $z \parallel [001]$. Consequently, a suppressed DSOI is expected for conduction band electrons in a [001]-oriented NW of, for instance, rectangular cross-section if the sides of the rectangle are parallel to $[110]$ and $[\bar{1}10]$. Remarkably, holes (unfilled valence band states) in Ge and Si NWs of such a geometry can feature an exceptionally strong Rashba-type SOI \cite{kloeffel:prb18}. 

\begin{table}[tb]
\caption{Terms proportional to $k_z^3$ in the Dresselhaus Hamiltonian $H_D$ [Eq.~(\ref{eq:HDSOIwithoutBfields})] for all configurations considered in Table~\ref{tab:EffDSOI}. For details, see Appendix~\ref{app:DerivationEffDresselhaus}. } 
\label{tab:CubicTermsEffDSOI}
\begin{tabular}{|c|c|c|c|}
\hline
$x$ & $y$ & $z$ & \mbox{ $k_z$-cubic terms in $H_D$ } \\
\hline 
$[100]$ & $[010]$ & $[001]$ & $0$ \\
$[110]$ & $[\bar{1}10]$ & $[001]$ & $0$ \\
$[001]$ & $[1\bar{1}0]$ & $[110]$ & $ \frac{b_D}{2} \sigma_y k_z^3 $ \\
$[\bar{1}10]$ & $[001]$ & $[110]$ & $ - \frac{b_D}{2} \sigma_x k_z^3 $ \\ 
$[\bar{1}11]$ & $[1\bar{1}2]$ & $[110]$ & $\frac{b_D}{2 \sqrt{3}} \left( \sigma_y - \sqrt{2} \sigma_x \right) k_z^3$ \\ 
$[11\bar{2}]$ & $[\bar{1}10]$ & $[111]$ & $0$ \\
$[1\bar{1}0]$ & $[11\bar{2}]$ & $[111]$ & $0$ \\
$[\bar{1}10]$ & $[111]$ & $[11\bar{2}]$ & $\frac{b_D}{2 \sqrt{3}} \sigma_x k_z^3$ \\ 
$[111]$ & $[1\bar{1}0]$ & $[11\bar{2}]$ & $- \frac{b_D}{2 \sqrt{3}} \sigma_y k_z^3$ \\
\hline
\end{tabular}
\end{table}

Since we are particularly interested in the regime of small~$k_z$, the terms proportional to $k_z^3$ (if present) in $H_D$ are not included in the effective DSOI Hamiltonian $H_{D, {\rm eff}}$. However, these $k_z$-cubic terms may be of high relevance to other research projects. In Table~\ref{tab:CubicTermsEffDSOI}, we therefore provide the omitted terms proportional to $k_z^3$ for all configurations listed in Table~\ref{tab:EffDSOI}.

\section{Convenient choice of axes}
\label{app:ChoiceOfAxes} 

In this appendix, we focus on the terms of type $\langle k_x k_y \rangle k_z$ in $H_{D, {\rm eff}}$ and discuss how the expectation value $\langle k_x k_y \rangle$ depends on the choice of the axes $x$ and $y$. As a simple example, we consider a NW whose cross-section is rectangular. The sides of the rectangle have the lengths $L_{\widetilde{x}}$ and $L_{\widetilde{y}}$ and are parallel to the axes $\widetilde{x}$ and $\widetilde{y}$, respectively. If hard-wall confinement is assumed and the origin of the coordinate system is at the center of the cross-section, the ground-state wave function of an electron in the NW has the orbital part
\begin{equation}
\psi_{\perp,g} = \frac{2}{\sqrt{L_{\widetilde{x}} L_{\widetilde{y}}}} \sin\Biggl[ \pi \left( \frac{\widetilde{x}}{L_{\widetilde{x}}} + \frac{1}{2} \right) \Biggr] \sin\Biggl[ \pi \left( \frac{\widetilde{y}}{L_{\widetilde{y}}} + \frac{1}{2} \right) \Biggr] 
\end{equation}       
for the transverse directions, provided that $|\widetilde{x}| < L_{\widetilde{x}}/2$ and $|\widetilde{y}| < L_{\widetilde{y}}/2$. This function for the orbital part may now be used to calculate $\langle k_x k_y \rangle = \bra{\psi_{\perp,g}} k_x k_y \ket{\psi_{\perp,g}}$. Let us first examine the case where the axes $x$ and $y$ are chosen, for instance, such that the momentum operators satisfy $k_x = ( k_{\widetilde{x}} + k_{\widetilde{y}} ) / \sqrt{2} $ and $k_y = ( k_{\widetilde{y}} - k_{\widetilde{x}} ) / \sqrt{2} $, i.e., the axes $x$ and $y$ are rotated with respect to the axes $\widetilde{x}$ and $\widetilde{y}$ by an angle of $\pi/4$. In this case, one finds $\langle k_x k_y \rangle = ( \langle k_{\widetilde{y}}^2 \rangle - \langle k_{\widetilde{x}}^2 \rangle ) / 2 \propto L_{\widetilde{y}}^{-2} - L_{\widetilde{x}}^{-2}$ with $k_{\widetilde{x}} = - i \partial_{\widetilde{x}}$ and  $k_{\widetilde{y}} = - i \partial_{\widetilde{y}}$. Thus $\langle k_x k_y \rangle$ is nonzero for $L_{\widetilde{x}} \neq L_{\widetilde{y}}$. In stark contrast, $\langle k_x k_y \rangle = 0$ even for $L_{\widetilde{x}} \neq L_{\widetilde{y}}$ if one chooses $x \parallel \widetilde{x}$ and $y \parallel \widetilde{y}$. As evident from this simple example of a rectangular cross-section, it is usually possible to choose the axes such that $\langle k_x k_y \rangle = 0$, which is why terms of type $\langle k_x k_y \rangle k_z$ were omitted in Table~\ref{tab:EffDSOI}. 

Figure~\ref{fig:sketch} shows a sketch of the sectorial annular cross-section considered in the main text and illustrates that the $y$~axis corresponds to a mirror axis. This choice is convenient for several reasons. In particular, our numerical calculations of the integrals confirm that $\langle k_x k_y \rangle = 0$ for all NWs analyzed in the main text. Consequently, Eqs.~(\ref{eqApp:HDeff11m2theta0kxky0}) and (\ref{eq:HDeff11m2MainText}) apply.

\section{Artifacts due to hard-wall confinement}
\label{app:ArtifactHardWall} 

Since the electrons are trapped inside the NWs, the expectation values $\langle k_x \rangle$, $\langle k_y \rangle$, $\langle k_x^3 \rangle$, $\langle k_y^3 \rangle$, $\langle k_x k_y^2 \rangle$, and $\langle k_x^2 k_y \rangle$ must be zero. By calculating these expectation values numerically, we find that $\langle k_x \rangle$, $\langle k_y \rangle$, $\langle k_x^3 \rangle$, and $\langle k_x k_y^2 \rangle$ indeed vanish for the NW geometries in the main text (Fig.~\ref{fig:sketch}). However, the evaluation of $\langle k_y^3 \rangle$ and $\langle k_x^2 k_y \rangle$ yields imaginary values. These unphysical results are artifacts of the hard-wall boundary conditions, as explained below. 

For the sake of simplicity, let us consider a test function $h(y)$ that fulfills the boundary conditions $h\left(y \leq y_1\right) = 0$ and $h\left(y \geq y_2\right) = 0$ imposed by hard-wall confinement. We note that $y_1$, $y_2$, and $h(y)$ may also depend on the coordinate $x$, which we omit in the notation for brevity. The function $h(y)$ is continuous over the entire range of $y$. In the range $y_1 < y < y_2$, the derivatives $\partial_y h(y)$, $\partial_y^2 h(y)$, and $\partial_y^3 h(y)$ exist and are continuous. In agreement with the properties of our functions~$\psi_\perp$, see Sec.~\ref{secsub:NonzeroInnerRadius} of the main text for details, we also assume that $h(y)$ is real-valued. Given this test function $h(y)$ and the position-space representation $k_y = - i \partial_y$, we first study integrals that are relevant for the calculation of $\langle k_y \rangle$. Integration by parts yields 
\begin{eqnarray}
\int_{y_1}^{y_2} dy \hspace{0.06cm} h(y) \partial_y h(y) &=& \left[h(y_2)\right]^2 - \left[h(y_1)\right]^2 - \int_{y_1}^{y_2} dy \hspace{0.06cm} [\partial_y h(y)] h(y) \nonumber \\
&=& - \int_{y_1}^{y_2} dy \hspace{0.06cm} h(y) \partial_y h(y) = 0 ,
\end{eqnarray}
from which one can conclude that $\langle k_y \rangle$ must vanish, in agreement with our numerical calculations. Next, we focus on integrals that are relevant for $\langle k_y^3 \rangle$ and find 
\begin{equation}
\int_{y_1}^{y_2} dy \hspace{0.06cm} h(y) \partial_y^3 h(y) 
= S - \int_{y_1}^{y_2} dy \hspace{0.06cm} \left[\partial_y^3 h(y)\right] h(y) , 
\label{eqApp:IntegralWithPartialyCubedNr1}
\end{equation} 
where 
\begin{equation}
S = \lim_{\epsilon \to 0^+} \left( \left. \left[ \partial_y h(y) \right]^2 \right|_{y = y_1 + \epsilon} - \left. \left[ \partial_y h(y) \right]^2 \right|_{y = y_2 - \epsilon} \right) . 
\label{eqApp:Sforky3}
\end{equation} 
The right-hand side of Eq.~(\ref{eqApp:IntegralWithPartialyCubedNr1}) is obtained from the left-hand side by performing three partial integrations and using again $h(y_1) = h(y_2) = 0$. The limit in the expression for $S$, see Eq.~(\ref{eqApp:Sforky3}), is needed since $\partial_y h(y)$ is not necessarily continuous at $y_1$ and $y_2$. Such discontinuities of the derivative do not occur in realistic wave functions and are a special feature caused by the hard-wall confinement. We note that Eq.~(\ref{eqApp:IntegralWithPartialyCubedNr1}) is equivalent to 
\begin{equation}
\int_{y_1}^{y_2} dy \hspace{0.06cm} h(y) \partial_y^3 h(y) = \frac{S}{2} .
\label{eqApp:IntegralWithPartialyCubedNr2}
\end{equation}
The quantity $S$ is real-valued. Moreover, $S$ is nonzero unless $|\partial_y h(y)|$ with $y_1 < y < y_2$ converges to the same value for $y \to y_1$ and $y \to y_2$. Consequently, the combination of Eq.~(\ref{eqApp:IntegralWithPartialyCubedNr2}) and $k_y^3 = i \partial_y^3$ implies that the hard-wall boundary conditions in our model allow for unphysical, imaginary results when $\langle k_y^3 \rangle$ is calculated, which is consistent with our numerical evaluation of $\langle k_y^3 \rangle$. In a similar way, one can explain that our imaginary results for $\langle k_x^2 k_y \rangle$, which would suggest that $k_x^2 k_y$ is not Hermitian, are artifacts caused by the considered hard-wall potential. Even in the case of hard-wall boundary conditions, however, the wave functions are always continuous (in contrast to their derivatives). Therefore, it turns out that the artifacts discussed in this appendix cannot occur in our calculations of $\langle k_x \rangle$, $\langle k_y \rangle$, $\langle k_x k_y \rangle$, $\langle k_x^2 \rangle$, and $\langle k_y^2 \rangle$.

\section{Size dependence}
\label{app:SizeDependence} 

The main purpose of the present appendix is to provide a detailed answer to the question how quantities in our calculations scale with the size of the NW cross-section. We therefore introduce the dimensionless parameters
\begin{gather}
\xi = k_\perp R_o , \\
r = \frac{R_i}{R_o} .
\end{gather}
The latter is simply the ratio of inner to outer radius. With these definitions, the determinant equation of Eq.~(\ref{eq:DeterminantEquatForkperp}) in the main text reads 
\begin{equation}
J_\alpha (r \xi) Y_\alpha (\xi) - J_\alpha (\xi) Y_\alpha (r \xi) = 0 . 
\label{eqApp:DeterminantEquatForXi}
\end{equation} 
We recall that $\alpha = m \pi / \phi_{\rm tot}$ depends on the angle $\phi_{\rm tot} = \phi_E - \phi_S$ and the considered value of $m \in \{ 1, 2, 3, \ldots \}$. Given $\alpha$ and $r$, one can use Eq.~(\ref{eqApp:DeterminantEquatForXi}) to find suitable numbers $\xi > 0$ and $C$ for which  
\begin{equation}
\psi_{\perp} = \mathscr{N}_{\perp} \left[ J_\alpha (\xi \mu) + C Y_\alpha (\xi \mu) \right] \sin\bigl[ \alpha \left( \phi - \phi_S \right) \bigr] 
\label{eqApp:psiperpAnsatzSACWithXi}
\end{equation}
is normalizable and fulfills all boundary conditions. The dimensionless coordinate 
\begin{equation} 
\mu = \frac{\rho}{R_o}
\end{equation} 
was introduced in Eq.~(\ref{eqApp:psiperpAnsatzSACWithXi}) for convenience and will prove very useful for rewriting our integrals. 

It is important to note that the numbers $\xi$ and $C$ for which $\psi_{\perp}$ vanishes at both $\rho = R_i$ and $\rho = R_o$ (i.e., $\mu = r$ and $\mu = 1$) depend solely on $\alpha$ and $r$. As a consequence, $\xi$ and $C$ do not change when $R_i$ and $R_o$ are varied such that their ratio $r$ remains constant. The eigenenergies $E_{\perp}$ [Eq.~(\ref{eq:EperpMainText})], which correspond to the subband edges of the NW, read 
\begin{equation}
E_{\perp} = \frac{\hbar^2 k_{\perp}^2}{2m_{\rm{eff}}} = \frac{\hbar^2 \xi^2}{2m_{\rm{eff}} R_o^2} .
\end{equation}   
That is, for any given shape of the cross-section (both $r$ and $\phi_{\rm tot}$ fixed) the eigenenergies $E_{\perp}$ are inversely proportional to the effective electron mass and the area $R_o^2 (1 - r^2) \phi_{\rm tot} / 2$ of the SAC. 

Next, we consider the normalization condition. By treating $\psi_{\perp} / \mathscr{N}_{\perp}$ as real-valued (see Sec.~\ref{secsub:NonzeroInnerRadius} for the justification) and by using
\begin{equation}
\int_{\phi_S}^{\phi_E} d\phi \sin^2\bigl[ \alpha \left( \phi - \phi_S \right) \bigr] = \frac{\phi_{\rm tot}}{2} , 
\end{equation}  
Eq.~(\ref{eq:NormalizationCondMainText}) can be rewritten as 
\begin{eqnarray}
1 &=& \frac{ \left| \mathscr{N}_{\perp} \right|^2 R_o^2 \phi_{\rm tot}}{2}  
\label{eqApp:NormalizConditRewritten} \\ 
& & \times \int_{r}^{1} d\mu \hspace{0.05cm} \mu \left[ J_\alpha^2 (\xi \mu) + 2 C J_\alpha (\xi \mu) Y_\alpha (\xi \mu) + C^2 Y_\alpha^2 (\xi \mu) \right] . \nonumber 
\end{eqnarray} 
Since suitable numbers for $\xi$ and $C$ depend solely on $\alpha$ and $r$, it is evident from Eq.~(\ref{eqApp:NormalizConditRewritten}) that the corresponding normalization factors satisfy $\left| \mathscr{N}_{\perp} \right| \propto  R_o^{-1}$ if $r$, $\phi_{\rm tot}$, and $m$ are fixed. 

The expectation values $\langle k_x^2 \rangle$ and $\langle k_y^2 \rangle$, which are needed for the effective DSOI terms (see Table~\ref{tab:EffDSOI} and Appendix~\ref{app:DerivationEffDresselhaus}), can be calculated via 
\begin{equation}
\langle k_{x,y}^2 \rangle = - \int_{R_i}^{R_o} d\rho \hspace{0.05cm} \rho \int_{\phi_S}^{\phi_E} d\phi \hspace{0.05cm} \psi_{\perp , g}^* \partial_{x,y}^2 \psi_{\perp , g} ,   
\label{eqApp:kxy2IntegralGeneral}
\end{equation}
where the asterisk indicates the complex conjugation and $\psi_{\perp , g}$ (the subscript $g$ stands for the ground state) is the function $\psi_\perp$ whose associated eigenenergy $E_\perp$ is minimal. Here we use the position-space representation $k_{x,y} = - i \partial_{x,y}$, thus omitting orbital corrections from magnetic fields, if present. In order to analyze how $\langle k_x^2 \rangle$ and $\langle k_y^2 \rangle$ scale with the size of the SAC, we recall some useful relations between the Cartesian and cylindrical coordinates considered in our work. The Cartesian coordinates $x$ and $y$ are related to the cylindrical coordinates $\rho$ and $\phi$ through $x = \rho \cos\phi$ and $y = \rho \sin\phi$. Consequently, the operators for the partial derivatives with respect to $x$ and $y$ can be written as 
\begin{eqnarray}
\partial_x &=& \cos\phi \hspace{0.06cm} \partial_\rho - \frac{\sin\phi}{\rho} \partial_\phi , \\
\partial_y &=& \sin\phi \hspace{0.06cm} \partial_\rho + \frac{\cos\phi}{\rho} \partial_\phi .  
\end{eqnarray}
By making use of the trigonometric identity $2 \sin\phi \cos\phi = \sin(2\phi)$, one finds
\begin{eqnarray}
\partial_x^2 &=& \cos^2\phi \hspace{0.06cm} \partial_\rho^2 + \frac{\sin^2\phi}{\rho} \partial_\rho + \frac{\sin^2\phi}{\rho^2} \partial_\phi^2 \nonumber \\ 
& & + \frac{\sin(2\phi)}{\rho^2} \partial_\phi - \frac{\sin(2\phi)}{\rho} \partial_\rho \partial_\phi 
\label{eqApp:partialXsquared} 
\end{eqnarray}
and
\begin{eqnarray}
\partial_y^2 &=& \sin^2\phi \hspace{0.06cm} \partial_\rho^2 + \frac{\cos^2\phi}{\rho} \partial_\rho + \frac{\cos^2\phi}{\rho^2} \partial_\phi^2 \nonumber \\ 
& & - \frac{\sin(2\phi)}{\rho^2} \partial_\phi + \frac{\sin(2\phi)}{\rho} \partial_\rho \partial_\phi 
\label{eqApp:partialYsquared}
\end{eqnarray} 
for the second derivatives. It can easily be verified that the well-known relation 
\begin{equation}
\partial_x^2 + \partial_y^2 = \partial_\rho^2 + \frac{1}{\rho} \partial_\rho + \frac{1}{\rho^2} \partial_\phi^2 
\end{equation}  
is consistent with Eqs.~(\ref{eqApp:partialXsquared}) and (\ref{eqApp:partialYsquared}). Finally, by inserting Eqs.~(\ref{eqApp:psiperpAnsatzSACWithXi}) and (\ref{eqApp:partialXsquared}) into Eq.~(\ref{eqApp:kxy2IntegralGeneral}) and using other relations discussed in this appendix, we find that $\langle k_x^2 \rangle$ can be expressed in the form
\begin{eqnarray}
\langle k_x^2 \rangle &=& - \left| \mathscr{N}_{\perp} \right|^2 \int_{r}^{1} d\mu \hspace{0.05cm} \mu \int_{\phi_S}^{\phi_E} d\phi  \hspace{0.05cm}  w_\alpha (\xi \mu) \sin\delta \nonumber \\
& & \times \Biggl[ \xi^2 w_\alpha^{\prime \prime} (\xi \mu) \cos^2\phi  \sin\delta 
\label{eqApp:kx2IntegralRewritten} \\ 
& & \mbox{ }\mbox{ }\mbox{ } + \frac{\xi}{\mu} w_\alpha^{\prime} (\xi \mu) \left( \sin^2\phi \sin\delta - \alpha \sin(2\phi) \cos\delta \right) \nonumber \\
& & \mbox{ }\mbox{ }\mbox{ } + \frac{1}{\mu^2} w_\alpha (\xi \mu) \left( \alpha \sin(2\phi) \cos\delta - \alpha^2 \sin^2\phi \sin\delta \right) \Biggr] , 
\nonumber 
\end{eqnarray} 
 where
\begin{equation}
\delta = \alpha \left( \phi - \phi_S \right)
\end{equation}
and
\begin{equation}
w_\alpha (\xi \mu) = J_\alpha (\xi \mu) + C Y_\alpha (\xi \mu) 
\end{equation}
for brevity. The prime and double prime in $w_\alpha^{\prime} (\xi \mu)$ and $w_\alpha^{\prime \prime} (\xi \mu)$, respectively, indicate the first and second derivative of the function $w_\alpha (\xi \mu)$ with respect to its argument $\xi \mu = k_\perp \rho$. Since we focus on the ground state when calculating $\langle k_x^2 \rangle$, it turns out that the integral in Eq.~(\ref{eqApp:kx2IntegralRewritten}), excluding the pre\-factor $\left| \mathscr{N}_{\perp} \right|^2$, depends solely on $r$, $\phi_S$, and $\phi_E$. Consequently, one finds $\langle k_x^2 \rangle \propto R_o^{-2}$ (as expected) if $r$, $\phi_S$, and $\phi_E$ are fixed. The same conclusion applies to $\langle k_y^2 \rangle$ and may be verified by inserting Eqs.~(\ref{eqApp:psiperpAnsatzSACWithXi}) and (\ref{eqApp:partialYsquared}) into Eq.~(\ref{eqApp:kxy2IntegralGeneral}) and rewriting the expression similarly to Eq.~(\ref{eqApp:kx2IntegralRewritten}). For the results presented in the main text, we always set $\phi_E = \pi - \phi_S$. Hence the SAC is mirror-symmetric with respect to the $y$ axis and $\langle k_x k_y \rangle = 0$. Considering $\phi_E = \pi - \phi_S = (\pi + \phi_{\rm tot})/2$, we thus conclude that the EDPs defined in Eqs.~(\ref{eq:HDeff11m2MainText}) and (\ref{eq:HDeff001MainText}) are of the form $\beta_{\rm eff} = b_D f(\phi_{\rm tot} , r) R_o^{-2} $ and $\bar{\beta}_{\rm eff} = b_D \bar{f}(\phi_{\rm tot} , r) R_o^{-2}$, respectively, where $f(\phi_{\rm tot} , r)$ and $\bar{f}(\phi_{\rm tot} , r)$ are functions that depend only on $\phi_{\rm tot}$ and $r$. Section~\ref{secsub:ScalingProperties} contains a discussion of these properties. The bar in $\bar{\beta}_{\rm eff}$ and $\bar{f}$ serves here as a convenient short-hand notation for the coincident configuration, meaning that the axes $x$, $y$, and $z$ coincide with the main crystallographic directions. Without the bar, $\beta_{\rm eff}$ and $f$ are associated with the noncoincident configuration (Sec.~\ref{secsub:OrientationCrystalAxes}), corresponding to the recently grown NWs of Ref.~\cite{friedl:nl18}.       

As a last remark, we would like to mention that some expressions in this appendix can be simplified by setting $C = r = 0$ in the special case of $R_i = 0$. For related information, we refer to Sec.~\ref{secsub:NoInnerRadius}.


\begin{thebibliography}{}


\bibitem{loss:pra98}{D. Loss and D. P. DiVincenzo, Phys. Rev. A {\bf 57}, 120 (1998).} 

\bibitem{nadjperge:nat10}{S. Nadj-Perge, S. M. Frolov, E. P. A. M. Bakkers, and L. P. Kouwenhoven, Nature (London) {\bf 468}, 1084 (2010).}

\bibitem{petersson:nat12}{K. D. Petersson, L. W. McFaul, M. D. Schroer, M. Jung, J. M. Taylor, A. A. Houck, and J. R. Petta, Nature (London) {\bf 490}, 380 (2012).}

\bibitem{kloeffel:annurev13}{C. Kloeffel and D. Loss, Annu. Rev. Condens. Matter Phys. {\bf 4}, 51 (2013).} 

\bibitem{maurand:ncomm16}{R. Maurand, X. Jehl, D. Kotekar-Patil, A. Corna, H. Bohuslavskyi, R. Lavi\'{e}ville, L. Hutin, S. Barraud, M. Vinet, M. Sanquer, and S. De Franceschi, Nat. Commun. {\bf 7}, 13575 (2016).} 

\bibitem{wang:nlt19}{R. Wang, R. S. Deacon, J. Sun, J. Yao, C. M. Lieber, and K. Ishibashi, Nano Lett. {\bf 19}, 1052 (2019).} 

\bibitem{alicea:rpp12}{J. Alicea, Rep. Prog. Phys. {\bf 75}, 076501 (2012).} 

\bibitem{beenakker:annurev13}{C. W. J. Beenakker, Annu. Rev. Condens. Matter Phys. {\bf 4}, 113 (2013).} 

\bibitem{klinovaja:prl14}{J. Klinovaja and D. Loss, Phys. Rev. Lett. {\bf 112}, 246403 (2014).} 

\bibitem{klinovaja:prb14}{J. Klinovaja and D. Loss, Phys. Rev. B {\bf 90}, 045118 (2014).}  

\bibitem{lutchyn:nrm18}{R. M. Lutchyn, E. P. A. M. Bakkers, L. P. Kouwenhoven, P. Krogstrup, C. M. Marcus, and Y. Oreg, Nat. Rev. Mater. {\bf 3}, 52 (2018).} 

\bibitem{manchon:nmat15}{A. Manchon, H. C. Koo, J. Nitta, S. M. Frolov, and R. A. Duine, Nat. Mater. {\bf 14}, 871 (2015).} 

\bibitem{nadjperge:prl12}{S. Nadj-Perge, V. S. Pribiag, J. W. G. van den Berg, K. Zuo, S. R. Plissard, E. P. A. M. Bakkers, S. M. Frolov, and L. P. Kouwenhoven, Phys. Rev. Lett. {\bf 108}, 166801 (2012).} 

\bibitem{rashba:prl03}{E. I. Rashba and Al. L. Efros, Phys. Rev. Lett. {\bf 91}, 126405 (2003).} 

\bibitem{golovach:prb06edsr}{V. N. Golovach, M. Borhani, and D. Loss, Phys. Rev. B {\bf 74}, 165319 (2006).}

\bibitem{flindt:prl06}{C. Flindt, A. S. Sorensen, and K. Flensberg, Phys. Rev. Lett. {\bf 97}, 240501 (2006).} 

\bibitem{lutchyn:prl10}{R. M. Lutchyn, J. D. Sau, and S. Das Sarma, Phys. Rev. Lett. {\bf 105}, 077001 (2010).} 

\bibitem{oreg:prl10}{Y. Oreg, G. Refael, and F. von Oppen, Phys. Rev. Lett. {\bf 105}, 177002 (2010).}

\bibitem{bihlmayer:njp15}{G. Bihlmayer, O. Rader, and R. Winkler, New J. Phys. {\bf 17}, 050202 (2015).}

\bibitem{bychkov:jetp84}{Y. A. Bychkov and E. I. Rashba, JETP Lett. {\bf 39}, 78 (1984).}

\bibitem{bychkov:jpcssp84}{Y. A. Bychkov and E. I. Rashba, J. Phys. C Solid State Phys. {\bf 17}, 6039 (1984).} 

\bibitem{winkler:book}{R. Winkler, {\it Spin-Orbit Coupling Effects in Two-Dimensional Electron and Hole Systems} (Springer, Berlin, 2003).} 

\bibitem{nitta:prl97}{J. Nitta, T. Akazaki, H. Takayanagi, and T. Enoki, Phys. Rev. Lett. {\bf 78}, 1335 (1997).} 

\bibitem{engels:prb97}{G. Engels, J. Lange, T. Sch\"{a}pers, and H. L\"{u}th, Phys. Rev. B {\bf 55}, R1958(R) (1997).} 

\bibitem{liang:nl12}{D. Liang and X. P. A. Gao, Nano Lett. {\bf 12}, 3263 (2012).} 

\bibitem{weigele:arX18}{P. J. Weigele, D. C. Marinescu, F. Dettwiler, J. Fu, S. Mack, J. C. Egues, D. D. Awschalom, and D. M. Zumb\"{u}hl, arXiv:1801.05657.} 

\bibitem{dresselhaus:pr55}{G. Dresselhaus, Phys. Rev. {\bf 100}, 580 (1955).} 

\bibitem{hanson:rmp07}{R. Hanson, L. P. Kouwenhoven, J. R. Petta, S. Tarucha, and L. M. K. Vandersypen, Rev. Mod. Phys. {\bf 79}, 1217 (2007).} 

\bibitem{dyakonov:sps86}{M. I. Dyakonov and V. Y. Kachorovskii, Sov. Phys. Semicond. {\bf 20}, 110 (1986).} 

\bibitem{balocchi:prl11}{A. Balocchi, Q. H. Duong, P. Renucci, B. L. Liu, C. Fontaine, T. Amand, D. Lagarde, and X. Marie, Phys. Rev. Lett. {\bf 107}, 136604 (2011).} 

\bibitem{flatte:physics11}{M. E. Flatt\'{e}, Physics {\bf 4}, 73 (2011).}

\bibitem{luo:prb11}{J.-W. Luo, L. Zhang, and A. Zunger, Phys. Rev. B {\bf 84}, 121303(R) (2011).} 

\bibitem{ganichev:pssb14}{S. D. Ganichev and L. E. Golub, Phys. Status Solidi B {\bf 251}, 1801 (2014).} 

\bibitem{kammermeier:prb16}{M. Kammermeier, P. Wenk, J. Schliemann, S. Heedt, and T. Sch\"{a}pers, Phys. Rev. B {\bf 93}, 205306 (2016).} 

\bibitem{campos:prb18}{T. Campos, P. E. Faria Junior, M. Gmitra, G. M. Sipahi, and J. Fabian, Phys. Rev. B {\bf 97}, 245402 (2018).} 

\bibitem{schliemann:prl03}{J. Schliemann, J. C. Egues, and D. Loss, Phys. Rev. Lett. {\bf 90}, 146801 (2003).} 

\bibitem{wagner:apl64}{R. S. Wagner and W. C. Ellis, Appl. Phys. Lett. {\bf 4}, 89 (1964).}

\bibitem{guniat:cr19}{L. G\"{u}niat, P. Caroff, and A. Fontcuberta i Morral, Chem. Rev. {\bf 119}, 8958 (2019).} 

\bibitem{casse:apl10}{M. Cass\'{e}, K. Tachi, S. Thiele, and T. Ernst, Appl. Phys. Lett. {\bf 96}, 123506 (2010).} 

\bibitem{barraud:edl12}{S. Barraud, R. Coquand, M. Cass\'{e}, M. Koyama, J.-M. Hartmann, V. Maffini-Alvaro, C. Comboroure, C. Vizioz, F. Aussenac, O. Faynot, and T. Poiroux, IEEE Electron Device Lett. {\bf 33}, 1526 (2012).} 

\bibitem{wu:nl04}{Y. Wu, Y. Cui, L. Huynh, C. J. Barrelet, D. C. Bell, and C. M. Lieber, Nano Lett. {\bf 4}, 433 (2004).} 

\bibitem{fortuna:sst10}{S. A. Fortuna and X. Li, Semicond. Sci. Technol. {\bf 25}, 024005 (2010).} 

\bibitem{hauge:nl15}{H. I. T. Hauge, M. A. Verheijen, S. Conesa-Boj, T. Etzelstorfer, M. Watzinger, D. Kriegner, I. Zardo, C. Fasolato, F. Capitani, P. Postorino, S. K\"{o}lling, A. Li, S. Assali, J. Stangl, and E. P. A. M. Bakkers, Nano Lett. {\bf 15}, 5855 (2015).} 

\bibitem{takase:scirep17}{K. Takase, Y. Ashikawa, G. Zhang, K. Tateno, and S. Sasaki, Sci. Rep. {\bf 7}, 930 (2017).} 

\bibitem{coquand:ulisproc12}{R. Coquand, S. Barraud, M. Cass\'{e}, P. Leroux, C. Vizioz, C. Comboroure, P. Perreau, E. Ernst, M.-P. Samson, V. Maffini-Alvaro, C. Tabone, S. Barnola, D. Munteanu, G. Ghibaudo, S. Monfray, F. Boeuf, and T. Poiroux, Proc. 13th Int. Conf. on Ultimate Integration on Silicon (ULIS), 2012, pp. 37--40.} 

\bibitem{voisin:nlt16}{B. Voisin, R. Maurand, S. Barraud, M. Vinet, X. Jehl, M. Sanquer, J. Renard, and S. De Franceschi, Nano Lett. {\bf 16}, 88 (2016).} 

\bibitem{calahorra:scirep17}{Y. Calahorra, A. Kelrich, S. Cohen, and D. Ritter, Sci. Rep. {\bf 7}, 40891 (2017).}

\bibitem{zhang:prl12}{J. J. Zhang, G. Katsaros, F. Montalenti, D. Scopece, R. O. Rezaev, C. Mickel, B. Rellinghaus, L. Miglio, S. De Franceschi, A. Rastelli, and O. G. Schmidt, Phys. Rev. Lett. {\bf 109}, 085502 (2012).} 

\bibitem{watzinger:nlt16}{H. Watzinger, C. Kloeffel, L. Vukusic, M. D. Rossell, V. Sessi, J. Kukucka, R. Kirchschlager, E. Lausecker, A. Truhlar, M. Glaser, A. Rastelli, A. Fuhrer, D. Loss, and G. Katsaros, Nano Lett. {\bf 16}, 6879 (2016).} 

\bibitem{li:nl18}{Y. Li, S.-X. Li, F. Gao, H.-O. Li, G. Xu, K. Wang, D. Liu, G. Cao, M. Xiao, T. Wang, J.-J. Zhang, G.-C. Guo, and G.-P. Guo, Nano Lett. {\bf 18}, 2091 (2018).}

\bibitem{watzinger:ncomm18}{H. Watzinger, J. Kukucka, L. Vukusic, F. Gao, T. Wang, F. Sch\"{a}ffler, J.-J. Zhang, and G. Katsaros, Nat. Commun. {\bf 9}, 3902 (2018).} 

\bibitem{friedl:nl18}{M. Friedl, K. Cerveny, P. Weigele, G. T\"{u}t\"{u}nc\"{u}oglu, S. Marti-Sanchez, C. Huang, T. Patlatiuk, H. Potts, Z. Sun, M. O. Hill, L. G\"{u}niat, W. Kim, M. Zamani, V. G. Dubrovskii, J. Arbiol, L. J. Lauhon, D. M. Zumb\"{u}hl, and A. Fontcuberta i Morral, Nano Lett. {\bf 18}, 2666 (2018).} 

\bibitem{alicea:nphys11}{J. Alicea, Y. Oreg, G. Refael, F. von Oppen, and M. P. A. Fisher, Nat. Phys. {\bf 7}, 412 (2011).} 

\bibitem{bringer:prb19}{A. Bringer, S. Heedt, and T. Sch\"{a}pers, Phys. Rev. B {\bf 99}, 085437 (2019).} 

\bibitem{aseev:nl19}{P. Aseev, A. Fursina, F. Boekhout, F. Krizek, J. E. Sestoft, F. Borsoi, S. Heedt, G. Wang, L. Binci, S. Marti-Sanchez, T. Swoboda, R. Koops, E. Uccelli, J. Arbiol, P. Krogstrup, L. P. Kouwenhoven, and P. Caroff, Nano Lett. {\bf 19}, 218 (2019).}  

\bibitem{mourik:sci12}{V. Mourik, K. Zuo, S. M. Frolov, S. R. Plissard, E. P. A. M. Bakkers, and L. P. Kouwenhoven, Science {\bf 336}, 1003 (2012).} 

\bibitem{vandenberg:prl13}{J. W. G. van den Berg, S. Nadj-Perge, V. S. Pribiag, S. R. Plissard, E. P. A. M. Bakkers, S. M. Frolov, and L. P. Kouwenhoven, Phys. Rev. Lett. {\bf 110}, 066806 (2013).} 

\bibitem{vanweperen:prb15}{I. van Weperen, B. Tarasinski, D. Eeltink, V. S. Pribiag, S. R. Plissard, E. P. A. M. Bakkers, L. P. Kouwenhoven, and M. Wimmer, Phys. Rev. B {\bf 91}, 201413(R) (2015).} 

\bibitem{gmitra:prb16}{M. Gmitra and J. Fabian, Phys. Rev. B {\bf 94}, 165202 (2016).} 

\bibitem{csontos:prb09}{D. Csontos, P. Brusheim, U. Z\"{u}licke, and H. Q. Xu, Phys. Rev. B {\bf 79}, 155323 (2009).}

\bibitem{nowak:prb13}{M. P. Nowak and B. Szafran, Phys. Rev. B {\bf 87}, 205436 (2013).} 

\bibitem{jespersen:prb15}{T. S. Jespersen, J. R. Hauptmann, C. B. Sorensen, and J. Nygard, Phys. Rev. B {\bf 91}, 041302(R) (2015).} 

\bibitem{heedt:nanoscale15}{S. Heedt, I. Otto, K. Sladek, H. Hardtdegen, J. Schubert, N. Demarina, H. L\"{u}th, D. Gr\"{u}tzmacher, and T. Sch\"{a}pers, Nanoscale {\bf 7}, 18188 (2015).} 

\bibitem{degtyarev:scirep17}{V. E. Degtyarev, S. V. Khazanova, and N. V. Demarina, Sci. Rep. {\bf 7}, 3411 (2017).} 

\bibitem{footnote:bD}{The Dresselhaus coefficient $b_D$ introduced in Eq.~(\ref{eq:HDSOI}) is very often denoted by $\gamma$ in the literature (sometimes with added superscripts or subscripts). It corresponds to $b^{6c6c}_{41}$ in Ref.~\cite{winkler:book}.} 

\bibitem{kloeffel:prb11}{C. Kloeffel, M. Trif, and D. Loss, Phys. Rev. B {\bf 84}, 195314 (2011).}

\bibitem{kloeffel:prb18}{C. Kloeffel, M. J. Rancic, and D. Loss, Phys. Rev. B {\bf 97}, 235422 (2018).} 

\bibitem{streda:prl03}{P. Streda and P. Seba, Phys. Rev. Lett. {\bf 90}, 256601 (2003).}  

\bibitem{jusserand:prb95}{B. Jusserand, D. Richards, G. Allan, C. Priester, and B. Etienne, Phys. Rev. B {\bf 51}, 4707(R) (1995).}  

\bibitem{knap:prb96}{W. Knap, C. Skierbiszewski, A. Zduniak, E. Litwin-Staszewska, D. Bertho, F. Kobbi, J. L. Robert, G. E. Pikus, F. G. Pikus, S. V. Iordanskii, V. Mosser, K. Zekentes, and Y. B. Lyanda-Geller, Phys. Rev. B {\bf 53}, 3912 (1996).} 

\bibitem{miller:prl03}{J. B. Miller, D. M. Zumb\"{u}hl, C. M. Marcus, Y. B. Lyanda-Geller, D. Goldhaber-Gordon, K. Campman, and A. C. Gossard, Phys. Rev. Lett. {\bf 90}, 076807 (2003).} 

\bibitem{krich:prl07}{J. J. Krich and B. I. Halperin, Phys. Rev. Lett. {\bf 98}, 226802 (2007).} 

\bibitem{faniel:prb11}{S. Faniel, T. Matsuura, S. Mineshige, Y. Sekine, and T. Koga, Phys. Rev. B {\bf 83}, 115309 (2011).} 

\bibitem{walser:nphys12}{M. P. Walser, C. Reichl, W. Wegscheider, and G. Salis, Nat. Phys. {\bf 8}, 757 (2012).} 

\bibitem{walser:prb12}{M. P. Walser, U. Siegenthaler, V. Lechner, D. Schuh, S. D. Ganichev, W. Wegscheider, and G. Salis, Phys. Rev. B {\bf 86}, 195309 (2012).} 

\bibitem{dettwiler:prx17}{F. Dettwiler, J. Fu, S. Mack, P. J. Weigele, J. C. Egues, D. D. Awschalom, and D. M. Zumb\"{u}hl, Phys. Rev. X {\bf 7}, 031010 (2017).} 

\bibitem{stano:prb18}{P. Stano, C.-H. Hsu, M. Serina, L. C. Camenzind, D. M. Zumb\"{u}hl, and D. Loss, Phys. Rev. B {\bf 98}, 195314 (2018).} 

\bibitem{stano:prb19}{P. Stano, C.-H. Hsu, L. C. Camenzind, L. Yu, D. Zumb\"{u}hl, and D. Loss, Phys. Rev. B {\bf 99}, 085308 (2019).} 

\bibitem{marinescu:prl19}{D. C. Marinescu, P. J. Weigele, D. M. Zumb\"{u}hl, and J. C. Egues, Phys. Rev. Lett. {\bf 122}, 156601 (2019).} 

\bibitem{footnote:aR}{The Rashba coefficient $a_R$ corresponds to $r^{6c6c}_{41}$ in Ref.~\cite{winkler:book}.} 

\bibitem{devries:nl18}{F. K. de Vries, J. Shen, R. J. Skolasinski, M. P. Nowak, D. Varjas, L. Wang, M. Wimmer, J. Ridderbos, F. A. Zwanenburg, A. Li, S. Koelling, M. A. Verheijen, E. P. A. M. Bakkers, and L. P. Kouwenhoven, Nano Lett. {\bf 18}, 6483 (2018).}

\bibitem{birpikus:book}{G. L. Bir and G. E. Pikus, {\it Symmetry and Strain-Induced Effects in Semiconductors} (Wiley, New York, 1974).} 

\bibitem{adachi:book1992}{S. Adachi, {\it Physical Properties of III-V Semiconductor Compounds: InP, InAs, GaAs, GaP, InGaAs, and InGaAsP} (Wiley, New York, 1992).} 

\bibitem{adachi:book2005}{S. Adachi, {\it Properties of Group-IV, III-V and II-VI Semiconductors} (Wiley, Chichester, 2005).} 

\bibitem{ivchenko:prb96}{E. L. Ivchenko, A. Y. Kaminski, and U. R\"{o}ssler, Phys. Rev. B {\bf 54}, 5852 (1996).} 

\bibitem{vervoort:prb97}{L. Vervoort, R. Ferreira, and P. Voisin, Phys. Rev. B {\bf 56}, R12744(R) (1997).} 

\bibitem{guettler:prb98}{T. Guettler, A. L. C. Triques, L. Vervoort, R. Ferreira, P. Roussignol, P. Voisin, D. Rondi, and J. C. Harmand, Phys. Rev. B {\bf 58}, R10179(R) (1998).} 

\bibitem{vervoort:sst99}{L. Vervoort, R. Ferreira, and P. Voisin, Semicond. Sci. Technol. {\bf 14}, 227 (1999).} 

\bibitem{olesberg:prb01}{J. T. Olesberg, W. H. Lau, M. E. Flatt\'{e}, C. Yu, E. Altunkaya, E. M. Shaw, T. C. Hasenberg, and T. F. Boggess, Phys. Rev. B {\bf 64}, 201301(R) (2001).}  

\bibitem{hall:prb03}{K. C. Hall, K. G\"{u}ndogdu, E. Altunkaya, W. H. Lau, M. E. Flatt\'{e}, T. F. Boggess, J. J. Zinck, W. B. Barvosa-Carter, and S. L. Skeith, Phys. Rev. B {\bf 68}, 115311 (2003).}  

\bibitem{golub:prb04}{L. E. Golub and E. L. Ivchenko, Phys. Rev. B {\bf 69}, 115333 (2004).}  

\bibitem{nestoklon:prb08}{M. O. Nestoklon, E. L. Ivchenko, J.-M. Jancu, and P. Voisin, Phys. Rev. B {\bf 77}, 155328 (2008).}  

\bibitem{prada:njp11}{M. Prada, G. Klimeck, and R. Joynt, New J. Phys. {\bf 13}, 013009 (2011).} 

\bibitem{furthmeier:ncomm16}{S. Furthmeier, F. Dirnberger, M. Gmitra, A. Bayer, M. Forsch, J. Hubmann, C. Sch\"{u}ller, E. Reiger, J. Fabian, T. Korn, and D. Bougeard, Nat. Commun. {\bf 7}, 12413 (2016).}  

\bibitem{wojcik:apl19}{P. Wojcik, A. Bertoni, and G. Goldoni, Appl. Phys. Lett. {\bf 114}, 073102 (2019).} 



\end{thebibliography}
\end{document}